\documentclass[twocolumn]{aastex631}

\usepackage{amsmath}
\usepackage{multirow}
\usepackage{array}
\usepackage{booktabs}

\begin{document}

\title{Searching for quasi periodic oscillations in optical and $\gamma$-ray emissions $\&$ black hole mass estimation of blazar ON 246}

\author[0000-0002-5221-0822]{Ajay Sharma}
\altaffiliation{ajjjkhoj@gmail.com}
\affiliation{S. N. Bose National Centre for Basic Sciences, Block JD, Salt Lake, Kolkata 700106, India}

\author[0000-0002-1173-7310]{Raj Prince}
\altaffiliation{rajprince59.bhu@gmail.com}
\affiliation{Department of Physics, Institute of Science, Banaras Hindu University, Varanasi, Uttar Pradesh, 221005, India}

\author[0000-0003-1071-5854]{Debanjan Bose}
\altaffiliation{debanjan.tifr@gmail.com}
\affiliation{Department of Physics, Central University of Kashmir, Ganderbal, 191131, India}




\begin{abstract}
We report the detection of a potential quasi-periodic signal with a period of $\sim$2 years in the blazar ON 246, based on Fermi-LAT ($\gamma$-rays) and ASAS-SN (optical) observations spanning 11.5 years (MJD 55932–60081). We applied various techniques to investigate periodic signatures in the light curves, including the Lomb-Scargle periodogram (LSP), Weighted Wavelet Z-transform (WWZ), and REDFIT. The significance of the signals detected in LSP and WWZ was assessed using two independent approaches: Monte Carlo simulations and red noise modeling. Our analysis revealed a dominant peak in the $\gamma$-ray and optical light curves, with a significance level exceeding 3$\sigma$ in both LSP and WWZ, consistently persisting throughout the observation period. Additionally, the REDFIT analysis confirmed the presence of a quasi-periodic signal at $\sim$0.00134 $day^{-1}$ with a 99$\%$ confidence threshold. To explain the observed quasi-periodic variations in $\gamma$-ray and optical emissions, we explored various potential physical mechanisms. Our analysis suggests that the detected periodicity could originate from a supermassive binary black hole (SMBBH) system or the jet-induced orbital motion within such a system. Based on variability characteristics, we estimated the black hole mass of ON 246. The study suggests that the mass lies within the range of approximately $(0.142 - 8.22) \times 10^9 \ M_{\odot}$. 
\end{abstract}

\section{\textbf{Introduction}} \label{sec:intro}
Active galactic nuclei (AGNs) are one of the most energetic astrophysical objects in the universe, powered by the accretion of matter of galaxy on supermassive black hole (SMBHs) with a mass in the range $10^6 - 10^{10} \rm{M}_{\odot}$ \citep{soƚtan1982masses}. Blazars, a subset of radio-loud AGNs, are among the most luminous objects with the bolometic luminosity in the range of $10^{41} - 10^{48}$ erg s$^{-1}$. Blazars produce relativistic jets that are aligned within a few degrees ($<5^{\circ}$) of our line of sight \citep{ghisellini1993relativistic, urry1995unified, blandford2019relativistic} and emit radiation over entire electromagnetic (EM) spectrum, from radio to very high energy ($>$100 GeV) $\gamma$-rays \citep{urry1995unified, ulrich1997variability, padovani2017active}. Blazars are further classified into two subclasses: BL Lacertae (BL Lacs) and flat-spectrum radio quasars (FSRQs), based on the characteristics and strength of broad emission lines in their optical spectra. BL Lacs exhibit featureless nonthermal optical spectra (very weak or absence of the lines), while the FSRQs show bright and strong broad emission lines with equivalent width (EW)$> 5 \ \text{\AA}$ in the rest frame \citep{giommi2012simplified}.\par
Observational studies have shown that these sources exhibit rapid and large flux modulations across the entire band of the electromagnetic (EM) spectrum from radio to VHE $\gamma$-ray, jet-dominated nonthermal emission that leads to the double-humped spectral energy distribution \citep{ulrich1997variability, fossati1998unifying}. The variation in emission provides valuable insights into various aspects of blazars, including structures, underlying emission mechanisms/processes, and physical parameters of SMBHs \citep{ulrich1997variability, gupta2017multi}. The observed flux variability timescale in all bands ranges from minutes to several years. Blazar's central region is very compact and often difficult to resolve directly with current facilities. By analyzing the rapid variability with timescale from minutes to hours, one can constrain the emission-region sizes of these sources effectively, and utilizing the simultaneous observations with theoretical models, we can constrain the physical parameters of jets \citep{blandford1982reverberation, tavecchio1998constraints, li2018fast, pandey2022detection}.\par
Observations have shown that blazar's flux variations are stochastic, nonlinear, and aperiodic in nature and well characterized by the simplest model of Continuous Autoregressive Moving Average [CARMA(p,q)] \citep{kelly2009variations}, also known as red noise model, but a small percentage of sources in the entire blazar population exhibit regular variations in the light curves, such particular phenomenon is known as quasi-periodic oscillation (QPO), which appear to be rare in AGNs. Such kinds of regular variations have been observed across the entire EM spectrum with the diverse timescales, ranging from minutes through months to years \citep{urry1993multiwavelength, wagner1995intraday, petry2000multiwavelength, katarzynski2001multifrequency, aleksic2011magic, sandrinelli2014long, carnerero2017dissecting, sarkar2019long, raiteri2021dual, sobolewska2014stochastic, gupta2008multicolor, gupta2019characterizing, mao2024radio}. The observed diverse timescale of QPOs may be associated with different underlying physical mechanisms and radiation processes in blazars. Intraday variability with timescales ranging from minutes to several hours may be originated via rotating inhomogeneous helical jet or current-driven kink instabilities \citep{raiteri2021complex, raiteri2021dual, jorstad2022rapid}. A short-term variability with a timescale ranging from days to a few months is believed to originate from the helical motion of magnetized plasma blob within the jet or perturbation in accretion disk at the innermost stable circular orbit \citep{zhou201834, gupta2019characterizing, sarkar2021multiwaveband, roy2022transient, banerjee2023detection, prince2023quasi, sharma2024detection, tantry2025study}. A long-term variability of timescale from several months to years may be associated with supermassive binary black hole (SMBBHs; \cite{begelman1980massive}) systems or jet structures. This interpretation has been adopted in several studies, e.g.\citep{graham2015possible, sandrinelli2016quasi, gupta2017multi, zhang2017gamma, li2021detection, ren2021detection, ren2021radiodetection, haiyan2023detection, li2023quasi}. The two most promising candidates, PG 1553+113 and OJ 287, have been reported for hosting an SMBBH system \citep{sillanpaa1988oj, valtonen2011testing, ackermann2015multiwavelength, tavani2018blazar, adhikari2024constraining}. The $\gamma$-ray emission in blazars originates from their relativistic jets. Investigating quasi-periodic variations in $\gamma$-ray emissions not only deepens our understanding of jet physics but also provides insights into particle acceleration mechanisms and jet dynamics. This has become possible due to the continuous monitoring capability of Fermi’s Large Area Telescope (Fermi-LAT). Leveraging long-term observations from Fermi-LAT, numerous strong QPOs in the $\gamma$-ray band have been reported in the literature. Additionally, in recent decades, systematic searches have been performed for QPOs in different wavebands for a number of sources \citep{gierlinski2008periodicity, gupta2008multicolor, gupta2008periodic, lachowicz20094, king2013quasi, alston2015discovery, bhatta2016detection, pan2016detection, bhatta2020nature, penil2020systematic, bhatta2021characterizing, ren2021detection, yang2021gaussian, wang2022possible, gong2022quasiperiodic, gong2023two, otero2023multiwavelength, lu2024research, ren2024possible}. In recent years, some QPO studies with high statistical significance have been reported in different EM bands. For instance, \citet{tripathi2021quasi} reported in radio, \citet{roy2022detection} in optical, \citet{smith2023qpo}, and \citet{gupta2019detection} in $\gamma$-ray. 

Additionally, some reported QPOs have also been interpreted by other geometrical models, such as pulsating accretion flow instability, jet precession, and Lense-Thirring precession of accretion disks \citep{romero2000beaming, rieger2005helical, stella1997lense}. Apart from the long-term persistent QPOs, several transient QPOs have also been reported \citep{zhou201834, benkhali2020evaluating, penil2020systematic, das2023detection, prince2023quasi, ren2023quasi, sharma2024detection}. The physical mechanisms of transient QPOs have been attributed to the orbiting hotspots on the disks, or close to the innermost stable circular orbits, magnetic reconnection within the jet, and helical orbital motion of blobs in the jet under the influence of magnetic field \citep{zhang1990rotation, mangalam1993accretion, gupta2008multicolor, gupta2008periodic, gupta2019detection, huang2013magnetic, mohan2015kinematics}. Thus, QPO studies play a crucial role in understanding the origin of such variations, the underlying radiation mechanisms, and the physical properties of SMBH systems.\par

The mass of a black hole is one of the most fundamental parameters as it plays a key role in shaping its emission properties and evolutionary behavior. In literature, several methods have been proposed to estimate the black hole mass: (1) the reverberation mapping technique \citep{kaspi2000reverberation}, (2) single-epoch spectral measurements/broad line width technique \citep{vestergaard2002determining}, (3) the gas and stellar dynamics technique \citep{genzel1997nature}, and (4) the variability timescale technique \citep{fan1999central, cheng1999basic, fan2005basic, fan2009estimations,yang2010central, liu2015constraints} and reference therein. \cite{pei2022estimation} derived the black hole mass of the blazar ON 246 to be $\sim8.08 \times 10^7 \ M_{\odot}$ based on certain assumptions, including the variability timescale of $\sim$1 day and a low Doppler factor of 0.48.\par

The strong radio source S3 1227+25 \citep{pauliny1972nrao} at R.A.=187$^{\circ}$.560, decl.=25$^{\circ}$.298, also known as ON 246 \citep{dixon1968high}, was first identified as a BL Lac candidate based on the correlation study between the ROSAT all-Sky Survey and the Hamburg Quasar Survey \citep{bade1994new}. Several studies have been carried out to classify this source based on the synchrotron peak frequency ($\nu_{peak}$). The observed $\nu_{peak}$ values are $10^{14.11}$ Hz \citep{wu2009debeamed}, $10^{14.41\pm 0.13}$ Hz \citep{fan2016spectral}, and $10^{14.91}$ Hz \citep{ackermann2015multiwavelength}. These studies indicate that this source lies near the boundary between intermediate synchrotron peak blazar (IBL) and high synchrotron peak blazar (HBL) in the classification scheme. VERITAS \citep{acharyya2023veritas} has detected this source in very high energy (VHE) band during MJD 57158-57160 and the average integral flux of $\left (4.51 \pm 0.44\right)\times 10^{-11} \ \rm{cm^{-2}} \ \rm{s^{-1}}$ above 0.12 TeV was reported. This source is one of the IBL objects detected so far in the VHE band. \citet{kharb2008parsec} resolved a parsec-scale core with radio observations. For the scientific study, we adopted the redshift (z) = 0.325 from \citet{acharyya2023veritas} in this work.\par
The paper is structured as follows: Section \ref{sec:sec2} covers multi-wavelength observations and reduction techniques. In Section \ref{sec:sec3}, we explore quasi-periodicity analysis using different methodologies, including the Lomb-Scargle Periodogram (LSP), Weighted Wavelet Z-transform (WWZ), and REDFIT. Section \ref{sec:sec4} focuses on Gaussian process modeling with a damped random walk model. In Section \ref{sec:sec5}, we assess the significance of quasi-periodic signals (QPOs) using two independent approaches, including Monte Carlo simulations and red noise modeling of light curves. Section \ref{sec:sec6} presents the findings of our QPO study, while Section \ref{sec:sec7} presents an interpretation of the observed QPO across multiple bands and wraps up with a conclusion.

\section{Multi-wavelength Observations } \label{sec:sec2}
\subsection{Fermi-LAT observation}

The Fermi Gamma-ray Space Telescope, launched by NASA on June 11, 2008, onboard two instruments: the Large Area Telescope (LAT) and the Gamma-ray Burst Monitor (GBM). Together, they enable comprehensive gamma-ray observations across a wide energy range, from a few keV to 500 GeV. The Fermi-LAT, a pair-conversion gamma-ray detector, is designed to explore high-energy gamma rays from $\sim$20 MeV to 500 GeV. It provides a wide field of view ($>$2 sr ), covering about 20$\%$ of the entire sky. Since its launch, Fermi-LAT has conducted all-sky surveys every three hours, providing near-continuous observations of $\gamma$-ray emissions from astrophysical sources \citep{atwood2009large}.\par
We collected Fermi-LAT data of blazar \textit{ON 246} during the period 2012 January 6 (MJD 55932) to 2023 May 17 (MET 60081) (over 1.6 years). During the data download procedure, we chose the energy range of 0.1-300 GeV with Pass8 class events (evclass==128, evtype==3) recommended by the Fermi-LAT collaboration from a region of interest (ROI) with a radius of $10^{\circ}$ centered at the source (R.A.=187$^{\circ}$.560, decl.=25$^{\circ}$.298 ). The analysis of $\gamma$-rays was performed following the standard procedures for point-source analysis using the Fermi Science Tools package (v11r05p3), provided by the Fermi Science Support Center. To minimize contamination from the Earth's limb, a zenith angle cut of $> 90^\circ$ was applied. The good time interval (GTI) data was extracted using the standard filtering using filter \texttt{$\text{(DATA\_QUAL > 0) \&\& (LAT\_CONFIG == 1)}$} to ensure high-quality observations. We used \texttt{GTLTCUBE} and \texttt{GTEXPOSURE} tools to calculate the integrated livetime as a function of sky position and off-axis angle and exposure, respectively. To model the galactic and extragalactic diffuse background emissions, we used models $\text{gll\_iem\_v07.fits}$\footnote{\label{fermi}\url{https://fermi.gsfc.nasa.gov/ssc/data/access/lat/BackgroundModels.html}} and $\text{iso\_P8R3\_SOURCE\_V3\_v1.txt}$\footref{fermi}, respectively. Further, we used the \textit{make4FGLxml.py} script to create the source model XML file, which contains the information about the source location and the best prediction of spectral form. The unbinned likelihood analysis was performed with \textit{GTLIKE} tool \citep{cash1979parameter, mattox1996likelihood} using the XML spectrum file, and the instrumental response function (IRF) "$\text{P8R3\_SOURCE\_V3}$" was adopted to get the final source spectrum. To find the significance of the source of interest, we used \textit{GTTSMAP} tool to calculate the test statistics (TS), which is defined as TS = 2$\Delta$log(likelihood) = -2log($\frac{L}{L_0}$), where $L$ and $L_0$ are the maximum likelihood of the model with and without a point source at the target location and maximum likelihood value fitted by the background model, respectively. The significance of finding the source at the specified position is accessed by the TS value with TS $\sim \sigma^2$ \citep{mattox1996likelihood}.\par
\begin{figure*}
    \centering
    \includegraphics[width=0.99\textwidth]{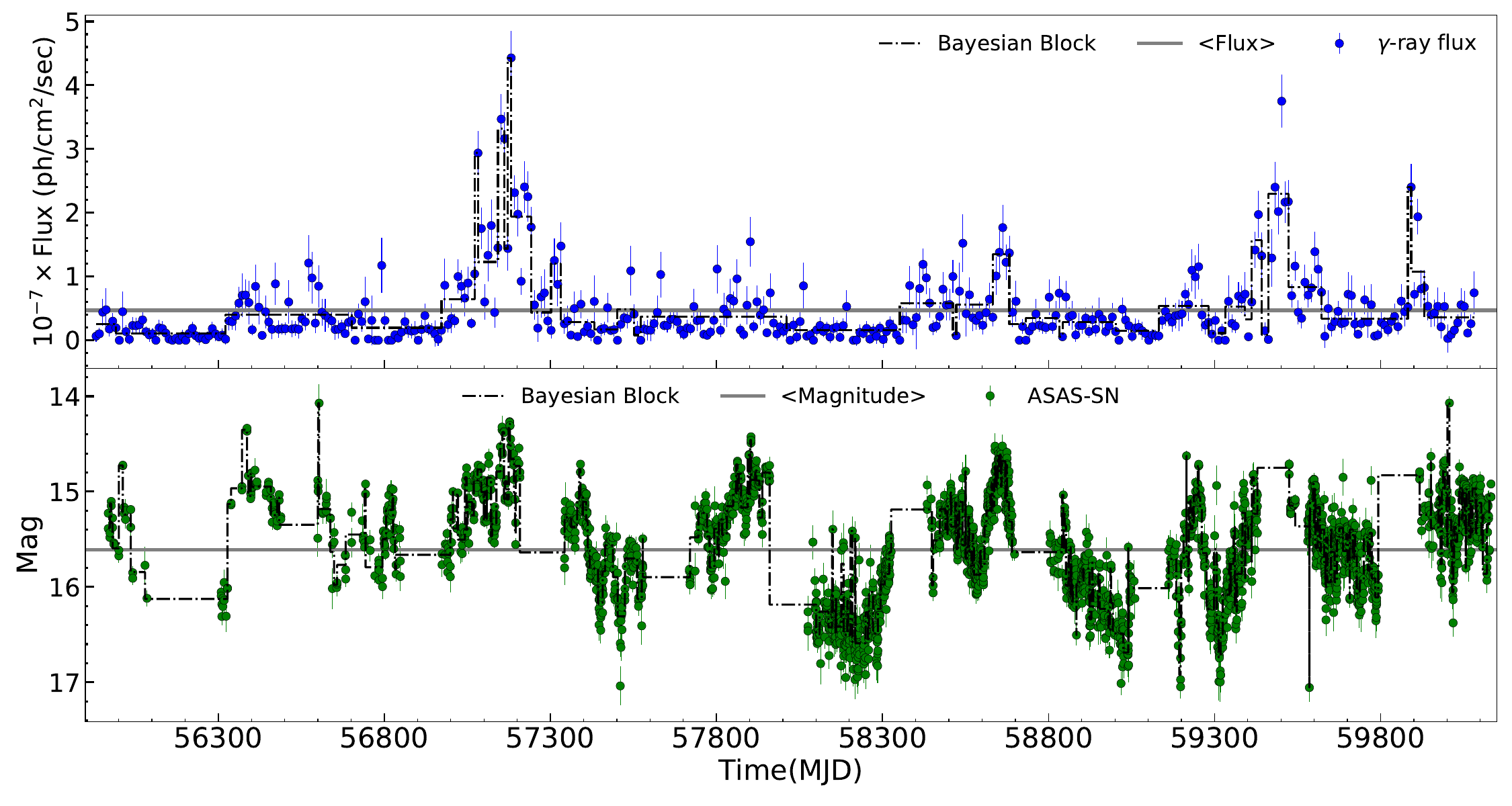}
    \caption{The figure presents the $\gamma$-ray and optical emissions observed between MJD 55900 and 60150. The top panel displays the weekly binned $\gamma$-ray flux (blue points) along with its Bayesian block (BB) representation (black curve). The bottom panel shows the ASAS-SN optical light curve (green) with the corresponding BB representation (black curve). The grey horizontal lines in both panels indicate the mean $\gamma$-ray flux and optical magnitude, respectively.}
    \label{Fig-MWL_lightcurve}    
\end{figure*}

We adopted a criterion with TS($\ge$9) for data points in the light curve, and a weekly binned light curve is generated using Fermipy\footnote{\url{https://fermipy.readthedocs.io/en/latest/}}. The resulting weekly binned $\gamma$-ray light curve is shown in Figure \ref{Fig-MWL_lightcurve}.

\subsection{\rm{ASAS-SN}}\label{ASAS}
All-Sky Automated Survey for Supernovae (ASAS-SN; \cite{shappee2014man, kochanek2017all}) is a global network of 24 telescopes that has been continuously scanning the extragalactic sky since 2012. ASAS-SN's limiting magnitude of V$\sim$16.5 - 17.5 and g$\sim$17.5 - 18.5 depending on lunation. ASAS-SN camera's field of view (FOV) is 4.5 deg$^2$, and the pixel scale and FWHM are $8^{''}.0$ and $\sim$2 pixels, respectively. For this study, we collected both bands' observations through the ASAS-SN Sky Patrol (V2.0\footnote{\url{http://asas-sn.ifa.hawaii.edu/skypatrol/}}; \citep{shappee2014man, hart2023asas}). 

\section{Periodicity Search}\label{sec:sec3}
We adopted various methodologies in search of a potential periodic signal in the $\gamma$-ray and optical light curves of blazar ON 246. Figure 1 illustrates the weekly binned $\gamma$-ray light curve along with optimal Bayesian Block representation (top panel) and an ASAS-SN optical light curve in the bottom panel.\par
The utilized methodologies include the Lomb-Scargle periodogram (LSP), Weighted Wavelet Z-Transform (WWZ), and a first-order autoregressive model $\left(AR(1)\right)$ in this QPO investigation. A detailed description and the observed findings from all the utilized methods are given in the following section \ref{LSP}, \ref{WWZ}.

\subsection{\rm{Lomb-Scargle Periodogram}}\label{LSP}
The Lomb-Scargle periodogram (LSP) \citep{lomb1976least, scargle1979studies} is one of the most widely used approaches in the literature to identify any potential periodic signal in a time series. In which a sinusoidal function fits the time series using the least square method. This approach is capable of handling the non-uniform sampling in the time series data efficiently by reducing the impact of noise and gaps and providing a precise measurement of the identified periodicity. In this study, we used the GLSP package to compute the Lomb-Scargle (LS) power. The expression of LS power is given as \citet{vanderplas2018understanding}:

\begin{equation}
\begin{split}
P_{LS}(f) = \frac{1}{2} \bigg[ &\frac{\left(\sum_{i=1}^{N} x_i \cos(2\pi f (t_i - \tau))\right)^2}{\sum_{i=1}^{N} \cos^2(2\pi f (t_i - \tau))} \\
&+ \frac{\left(\sum_{i=1}^N x_i \sin(2\pi f (t_i - \tau))\right)^2}{\sum_{i=1}^N \sin^2(2\pi f (t_i - \tau))} \bigg]
\end{split}
\end{equation}

where, $\tau$ is specified for each $f$ to ensure time-shift invariance:

\begin{equation}
    \tau = \frac{1}{4 \pi f} \tan^{-1} \left( \frac{\sum_{i=1}^N \sin\left( 4 \pi f t_i \right)}{\sum_{i=1}^N \cos\left( 4 \pi f t_i \right)} \right)
\end{equation}

where, we selected the minimum frequency $\left( f_{min} \right)$ and maximum frequency $\left( f_{max} \right)$ in temporal frequency range as 1/T and 1/2$\Delta T$, respectively, and here T and $\Delta T$ represent the total observation time frame and the time difference between two consecutive points in the light curve, respectively.\par
The LSP analysis reveals prominent peaks at frequencies of $\sim$0.00134 $\rm{day^{-1}}$ ($746\pm68$ days) in the $\gamma$-ray LSP (see Figure \ref{Fig-LSP_WWZ_Gamma}) and $\sim$0.00132 $\rm{day^{-1}}$ ($757\pm106$ days) in the optical LSP (see Figure \ref{Fig-LSP_WWZ_optical}). Both peaks have a local significance level exceeding 99.73$\%$. The uncertainty on the observed period is estimated by fitting a Gaussian function to the dominant LSP peak, and the obtained half-width and half maxima (HWHM) value is used as an uncertainty on period. The distribution of LS power as a function of frequency is given in Figure 2.  

\subsection{\rm{Weighted Wavelet Z-Transform}}\label{WWZ}
In contrast to the LSP approach, the Weighted Wavelet Z-transform (WWZ) \citep{foster1996wavelets} emerges as a powerful, robust, and widely used method in astronomical studies to identify any potential periodic pattern in irregularly sampled light curves. The WWZ method incorporates wavelet analysis,
enhancing the LSP’s capabilities by providing better localization of periodic signals in both temporal and spectral space. In studying the evolution of a periodic signal over time, this approach emerges as a powerful tool, enabling us to identify and characterize the nature of a periodic signal.\par
In this study, we adopted the abbreviated Morlet kernel that has the following functional form \citep{grossmann1984decomposition}:

\begin{equation}
    f[\omega (t - \tau)] = \exp[i \omega (t - \tau) - c \omega^2 (t - \tau)^2]
\end{equation}
and the corresponding WWZ map is given by,
\begin{equation}
    W[\omega, \tau: x(t)] = \omega^{1/2} \int x(t)f^* [\omega(t - \tau)] dt
\end{equation}

where, $f^*$ is the complex conjugate of the wavelet kernel f, $\omega$ is the frequency, and $\tau$ is the time-shift. This kernel acts as a windowed DFT, where the size of the window is determined by both the parameters $\omega$ and a constant \textit{c}. The resulting WWZ map offers a notable advantage; it not only identifies dominant periodicities but also provides insights into their duration over time.\par

\begin{figure*}
    \centering
   \includegraphics[scale=0.27]{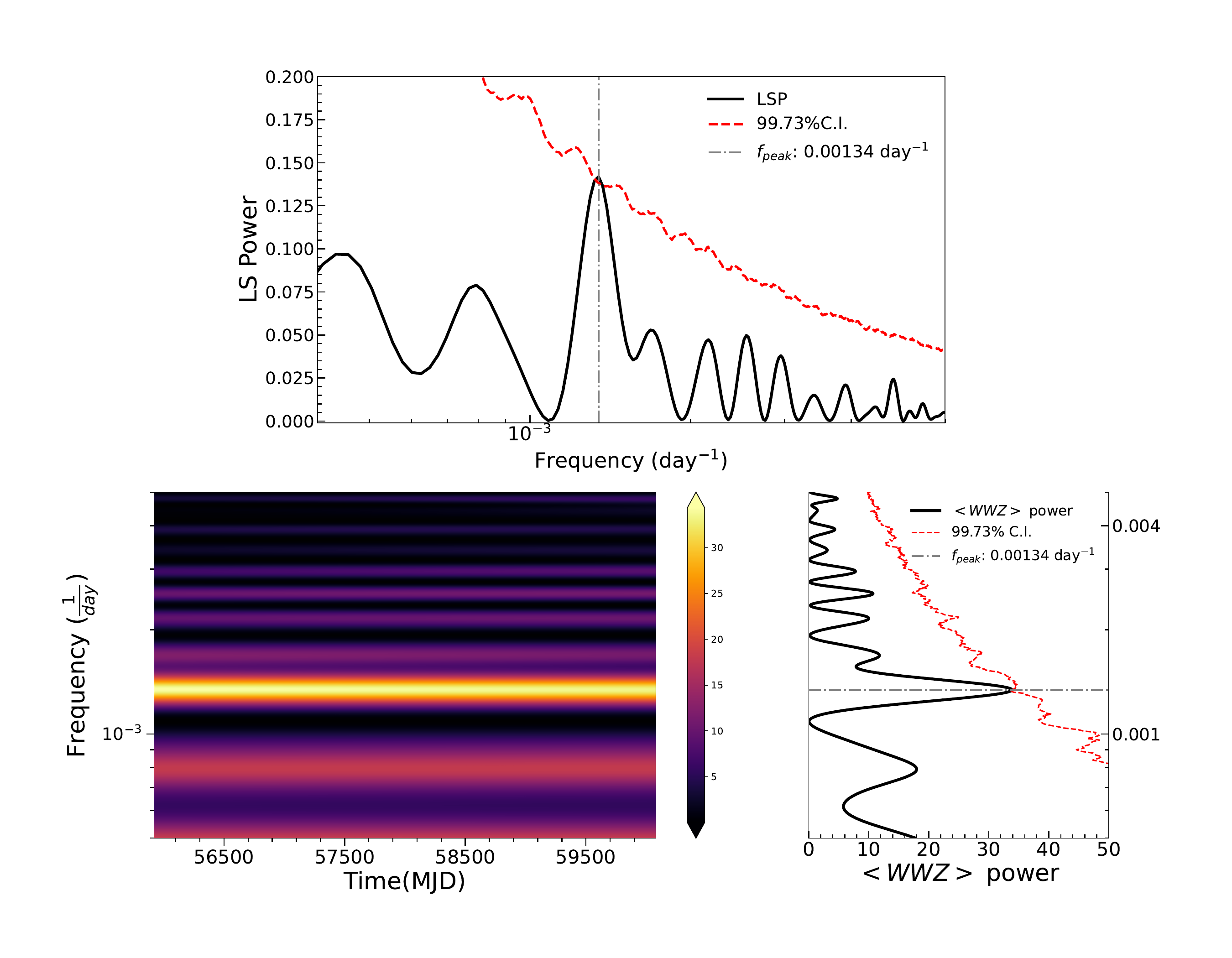}
    \caption{The $\gamma$-ray light curve is analyzed using the Lomb-Scargle Periodogram (LSP) and Weighted Wavelet Z-transform (WWZ) methods. The top panel shows the local significance of the detected peak at $\sim$0.00134 $day^{-1}$ in $\gamma$-ray LSP is exceeding 99.73$\%$.  The bottom panels display the WWZ map (left) and average wavelet power (right). The observed local significance of the detected peak at $\sim$0.00132 $day^{-1}$ in avg. wavelet has a significance level of $99.73\%$. }
    \label{Fig-LSP_WWZ_Gamma}    
\end{figure*}

\begin{figure*}
    \centering
    \includegraphics[scale=0.27]{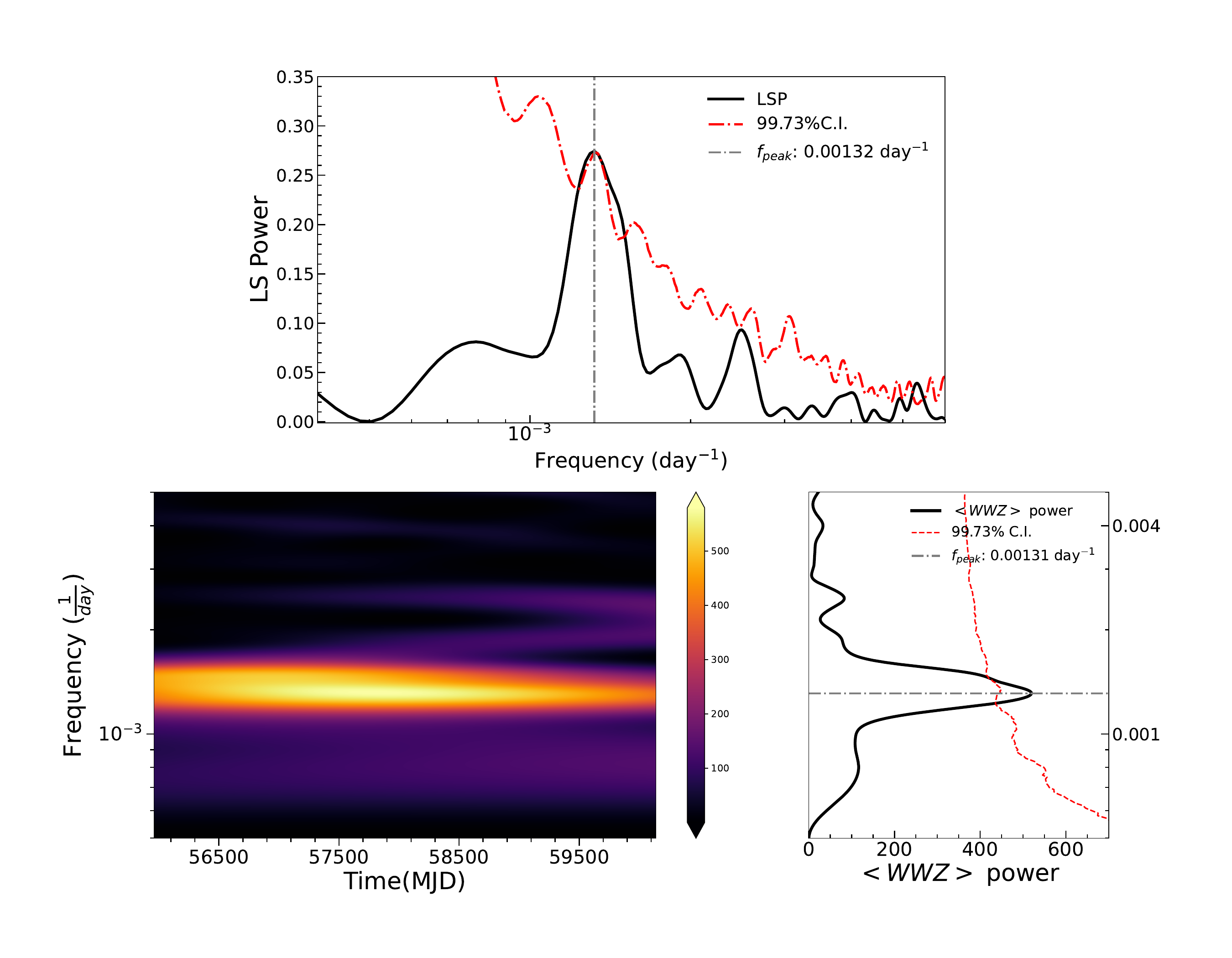}
    \caption{The detected QPO signals in the optical emissions from ON 246. The top panel shows the LSP with a dominant peak at $\sim$0.00132 $day^{-1}$ has a local significance level exceeding 99.73$\%$. The bottom panels display the wavelet map (bottom left panel) and avg. wavelet power at frequency of $\sim0.00131 \ day^{-1}$ with a significance level greater than 99.73$\%$. }
    \label{Fig-LSP_WWZ_optical}    
\end{figure*}

In this study, we used publicly available python code\footnote{\url{https://github.com/eaydin/WWZ}} \citep{m_emre_aydin_2017_375648} to generate the WWZ map. The observed power concentration is located around 0.00132 $day^{-1}$ ($757\pm80$ days) in the WWZ map utilizing $\gamma$-ray emissions (see Figure \ref{Fig-LSP_WWZ_Gamma}) and at $\sim$0.00131 $day^{-1}$ ($763\pm102$ days) in optical WWZ map (see Figure \ref{Fig-LSP_WWZ_optical}). In both cases, the observed local significance surpasses 99.73$\%$. The uncertainty on the period was estimated using the method as described in section \ref{LSP}.

\subsection{\rm{REDFIT}}\label{redfit}
The light curves of AGNs are typically unevenly sampled, of finite duration, and predominantly influenced by red noise, which arises from stochastic processes occurring in the accretion disk or jet. Red noise spectra are characteristic of autoregressive processes, where current activity is related to past behavior. The emissions from AGNs are effectively modeled using a first-order autoregressive (AR1) process. To model such behavior, the software programme \textsc{REDFIT}, developed by \citet{schulz2002redfit}, is specifically designed to analyze the stochastic nature of AGNs dominated by red noise. This software fits the light curve AR(1) process, where the current emission ($r_t$) depends linearly on the previous emission ($r_{t - 1}$) and a random error term ($\epsilon_t$). The AR(1) process is defined as:

\begin{equation}
    r(t_i)=A_i r(t_{i-1}) + \epsilon(t_i)
\end{equation}

where $r(t_i)$ is the flux value at time $t_i$ and $A_i = exp\left( \left[ \frac{t_{i-1} - t_i}{\tau}\right] \right) \in [0,1]$, A is the average autocorrelation coefficient computed from mean of the sampling intervals, $\tau$ is the time-scale of autoregressive process, and $\epsilon$ is a Gaussian-distributed random variable with zero mean and variance of unit. The power spectrum corresponding to the AR(1) process is given by 

\begin{equation}
    G_{rr}(f_i) = G_0 \frac{1 - A^2}{1 - 2 A cos\left( \frac{\pi f_i}{f_{Nyq}} \right) + A^2}
\end{equation}

where $G_0$ is the average spectral amplitude, $f_i$ are the frequencies, and $f_{Nyq}$ is representing the Nyquist frequency. \par
In our study, we used the publicly available 
\textsc{REDFIT}\footnote{\url{https://rdrr.io/cran/dplR/man/redfit.html}} code to analysis the light curve. In this method, the Nyquist frequency is defined as $f_{Nyq} = H_{fac}/ (2 \Delta t)$, where the factor $H_{fac}$ is introduced to prevent the noisy high-frequency end of the spectrum from influencing the fit, as described by Equation (7). The REDFIT analysis detected prominent peaks at frequencies of $\sim$0.00128 $\rm{day^{-1}}$ ($781\pm160$ days) in the $\gamma$-ray lightcurve (see the left panel of Figure \ref{Fig-REDFIT}) and $\sim$0.00132 $\rm{day^{-1}}$ ($757\pm160$ days) in the optical lightcurve (see the right panel of Figure \ref{Fig-REDFIT}). The uncertainties in the periods were estimated using the methodology described in Section~\ref{LSP}. 

\begin{figure}[hbt!]
    \centering
    \includegraphics[width=0.9\columnwidth]{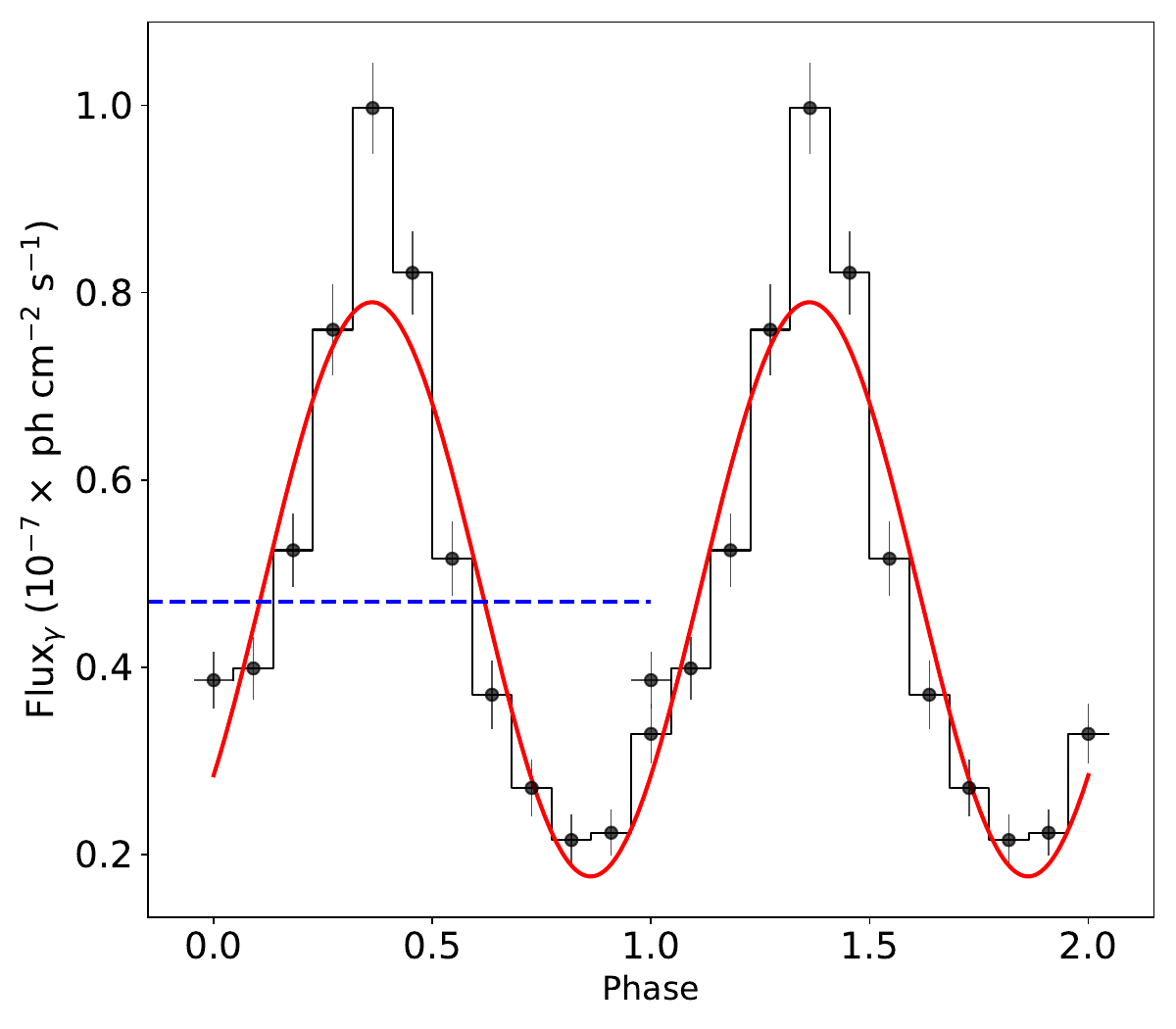} \vspace{4pt}
    
    \includegraphics[width=0.9\columnwidth]{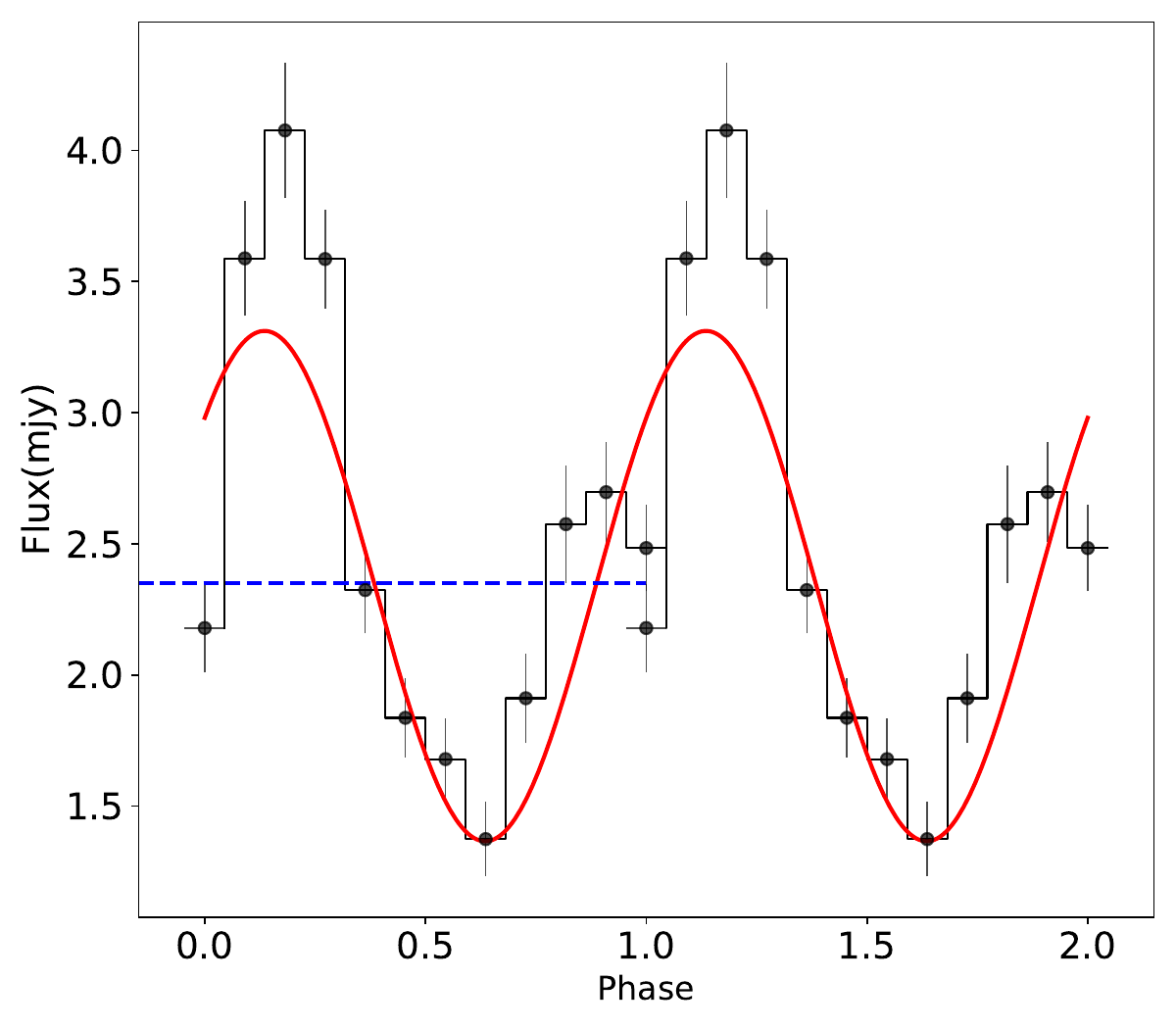}
    
    \caption{The folded Fermi-LAT and ASAS-SN light curves of ON 246 with a period of 746 and 757 days are shown in the top and bottom panels, respectively. The dashed blue line represents the mean value, and the sine functions (red) with frequencies of 0.00134 and 0.00132 day$^{-1}$ were fitted to the folded $\gamma$-ray and optical light curves, respectively. Two full period cycles are shown for better clarity.}
    \label{Fig-phase_folded}
\end{figure}

\begin{figure*}
    \centering
    \includegraphics[width=0.45\textwidth]{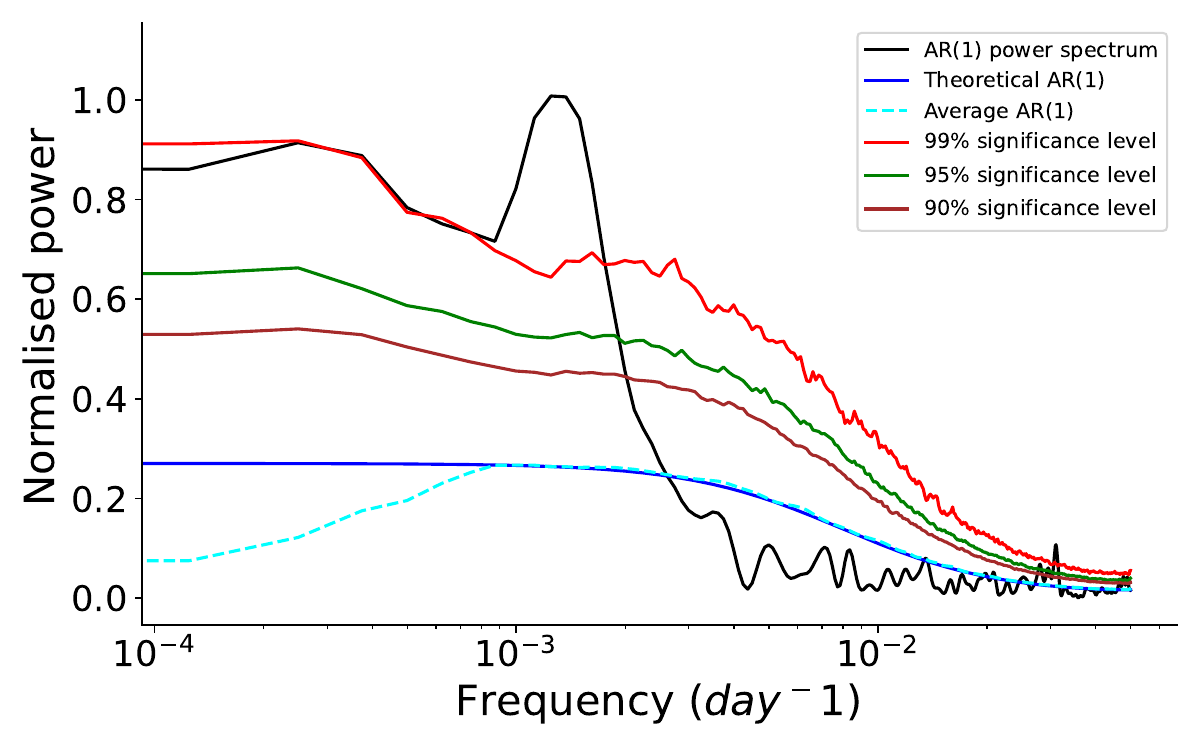} \hspace{1pt}
    \includegraphics[width=0.45\textwidth]{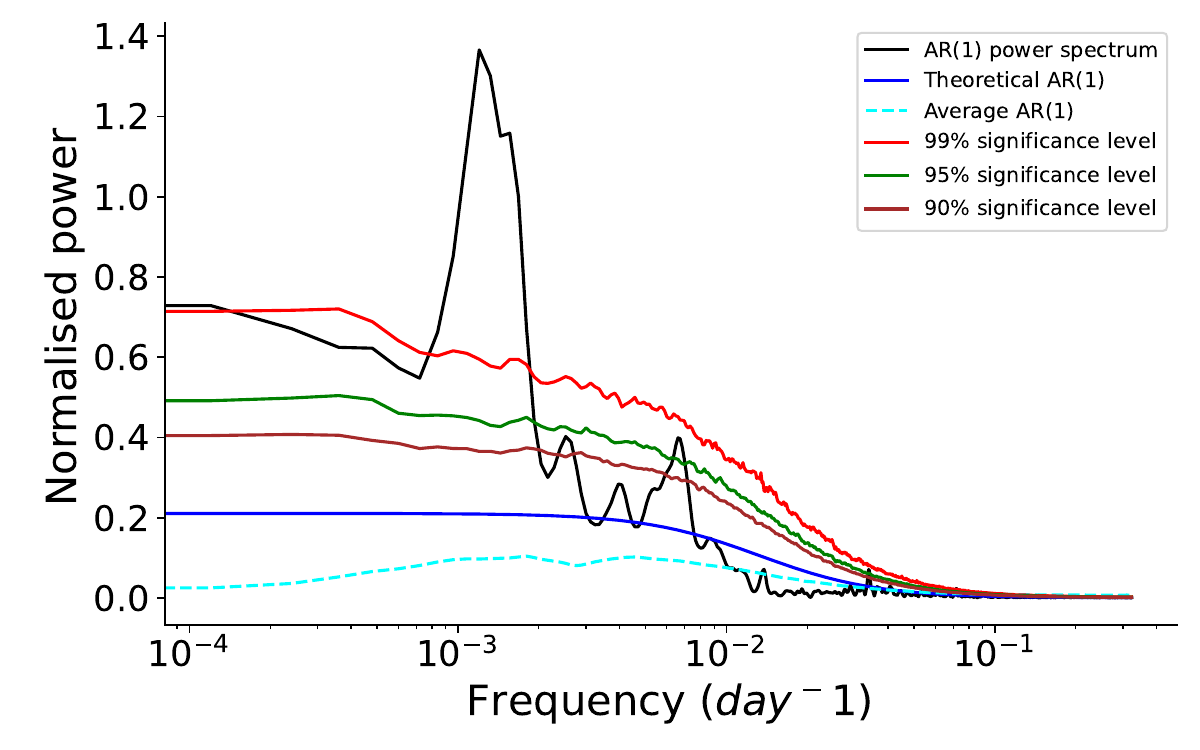}
    \caption{Analysis of the light curves, left panel represent the REDFIT curve of $\gamma$-ray emissions and right panel exhibit the REDFIT curve of optical emissions, using the AR(1) process with the REDFIT software. The red noise-corrected power spectrum (black) is presented alongside theoretical (blue) and average AR(1) (cyan) spectra. The significance levels of 99$\%$, 95$\%$, and 90$\%$ are indicated in red, green, and brown, respectively. }
    \label{Fig-REDFIT}    
\end{figure*}

\section{Gaussian Process modeling}\label{sec:sec4}
The observed variability in AGN is inherently stochastic. The AGN light curves can be well described by the stochastic processes, also known as Continuous Time Autoregressive Moving Average [CARMA(p, q)] processes \citep{kelly2014flexible}, defined as the solutions to the stochastic differential equation:

\begin{equation}
\begin{split}
\frac{d^p y(t)}{dt^p} + \alpha_{p-1}\frac{d^{p-1}y(t)}{dt^{p-1}}+...+\alpha_0 y(t) =\\
\beta_q \frac{d^q \epsilon(t)}{dt^q}+\beta_{q-1}\frac{d^{q-1}\epsilon(t)}{dt^{q-1}}+...+\beta_0 \epsilon(t),
\end{split}
\end{equation}

where, y(t) represents a time series, $\epsilon$(t) is a continuous time white noise process, and $\alpha_*$ and $\beta_*$ are the coefficients of autoregressive (AR) and moving average (MA) models, respectively. Here, p and q  are the order parameters of AR and MA models, respectively. \par
The simplest model is a continuous autoregressive [CAR(1)] model, also known as the Ornstein-Uhlenbeck process. It is a popular red noise model \citep{kelly2009variations, kozlowski2009quantifying, macleod2012description, ruan2012characterizing, zu2013quasar, moreno2019stochastic, burke2021characteristic, zhang2022characterizing, zhang2023gaussian, sharma2024probing, zhang2024discovering, sharma2024microquasars}, usually referred to as the Damped Random Walk (DRW) model, described by the following differential equation:

\begin{equation}
    \left[ \frac{d}{dt} + \frac{1}{\tau_{DRW}} \right] y(t) = \sigma_{DRW} \epsilon(t)
\end{equation}

where $\tau_{DRW}$  and $\sigma_{DRW}$ are the characteristic damping time-scale and amplitude of the DRW process, respectively. The mathematical form of the covariance function of the DRW model is defined as

\begin{equation}
\centering
    k(t_{nm}) = a \cdot \exp(-t_{nm} \, c),
\end{equation}

where $t_{nm} = | t_n -t_m|$ denotes the time lag between measurements m and n, with $a = 2 \sigma_{DRW}^2$ and $c = \frac{1}{\tau_{DRW}}$. The power spectral density (PSD) of the DRW model is defined as:
\begin{equation}
    S(\omega) = \sqrt{\frac{2}{\pi}} \frac{a}{c} \frac{1}{1 + (\frac{\omega}{c})^2}
\end{equation}

The DRW PSD  has a form of Broken Power Law (BPL), where the broken frequency $f_b$ corresponds to the characteristic damping timescale $\tau_{DRW} = \frac{1}{2\pi f_b}$.\par
In the best-fit parameters estimation of the DRW model for both light curves, we employed the Markov chain Monte Carlo (MCMC) algorithm provided by the \textsc{emcee}\footnote{\url{https://emcee.readthedocs.io/en/stable/}} package \citep{foreman2013emcee}. For the modeling, we employed the \textsc{EzTao}\footnote{\url{https://eztao.readthedocs.io/en/latest/index.html}} package, which is built on top of \textsc{celerite}\footnote{\url{https://celerite.readthedocs.io/en/stable/}}. In this study, we generated the distributions of the posterior parameters by running 10000 steps as burn-in and 20000 as burn-out.

\begin{figure*}
    \centering
    \includegraphics[width=0.9\textwidth]{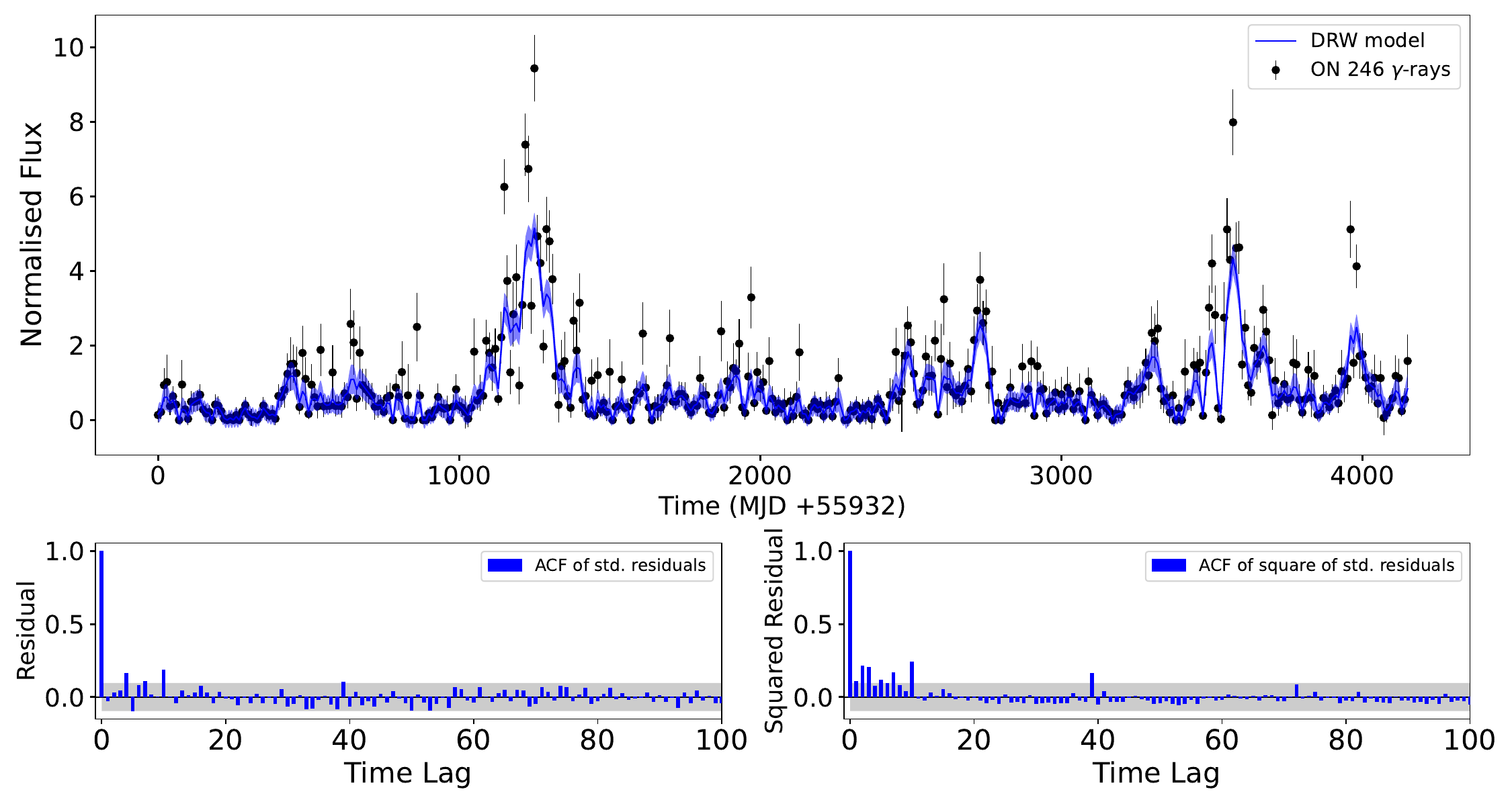} 
    \caption{The celerite modeling with DRW model was performed using the 10-day binned $\gamma$-ray light curve of blazar ON 246 over 4000 days from the time stamp MJD 55932. In this figure, the top panel depicts the $\gamma$-ray flux points with their uncertainties, along with the best-fitting profile of the DRW model in blue, including the 1$\sigma$ confidence interval. The bottom panels represent the autocorrelation functions (ACFs) of the standardized residuals (bottom left) and the squared of standardized residuals (bottom right), respectively, along with 95$\%$ confidence intervals of the white noise. }
    \label{Fig-DRW_gamm}    
\end{figure*}

\begin{figure*}
    \centering
    \includegraphics[width=0.9\textwidth]{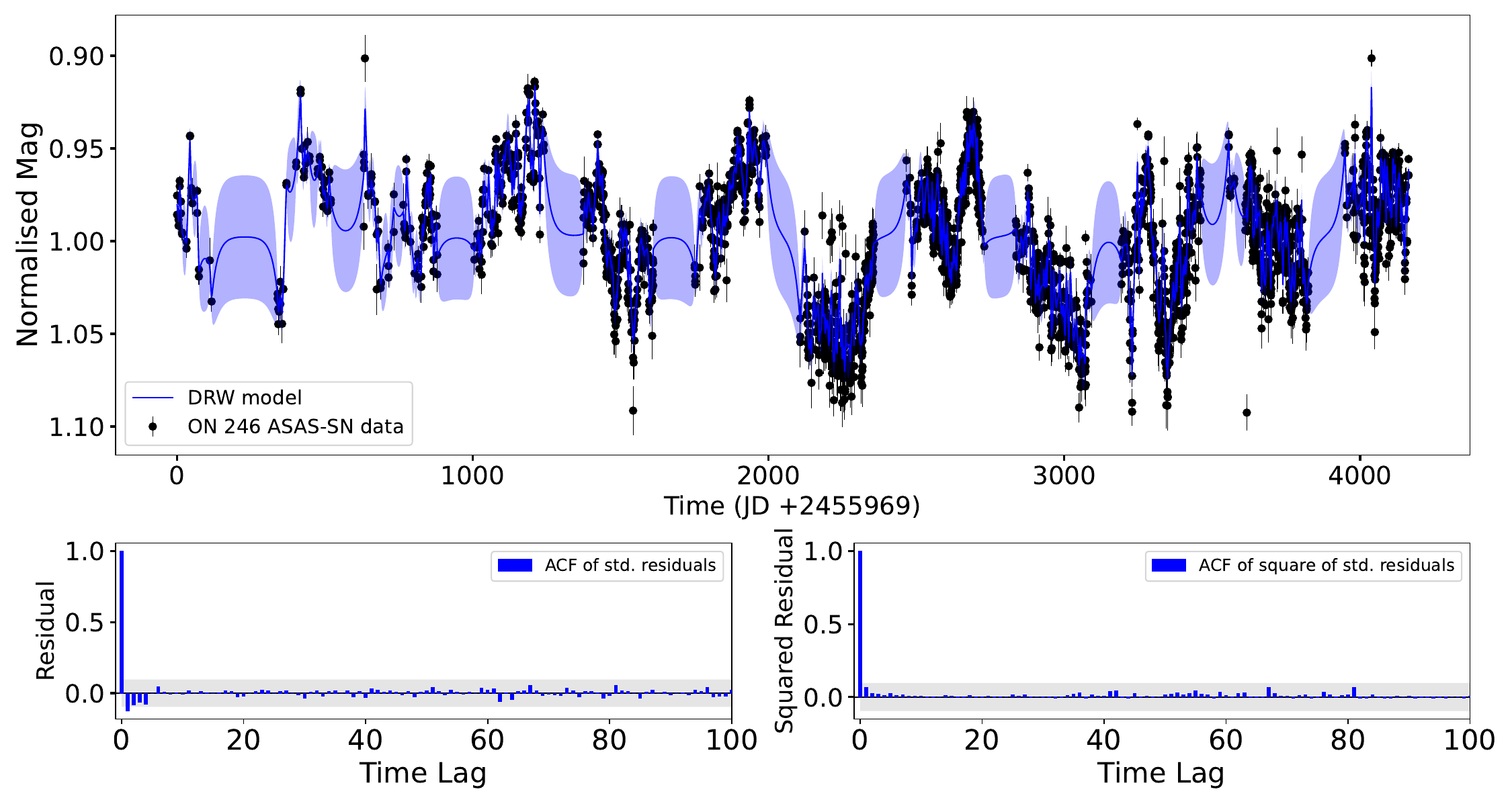} 
    \caption{This figure demonstrates the modeling of the ASAS-SN light curve of ON 246 with the DRW model. The top panel shows the best-fitting profile of the DRW modeling along with 1$\sigma$ confidence interval. The bottom panels represent the autocorrelation functions as described in Figure \ref{Fig-DRW_gamm}. }
    \label{Fig-DRW_optical}    
\end{figure*}

\begin{figure}[hbt!]
    \centering
    \includegraphics[width=0.45\textwidth]{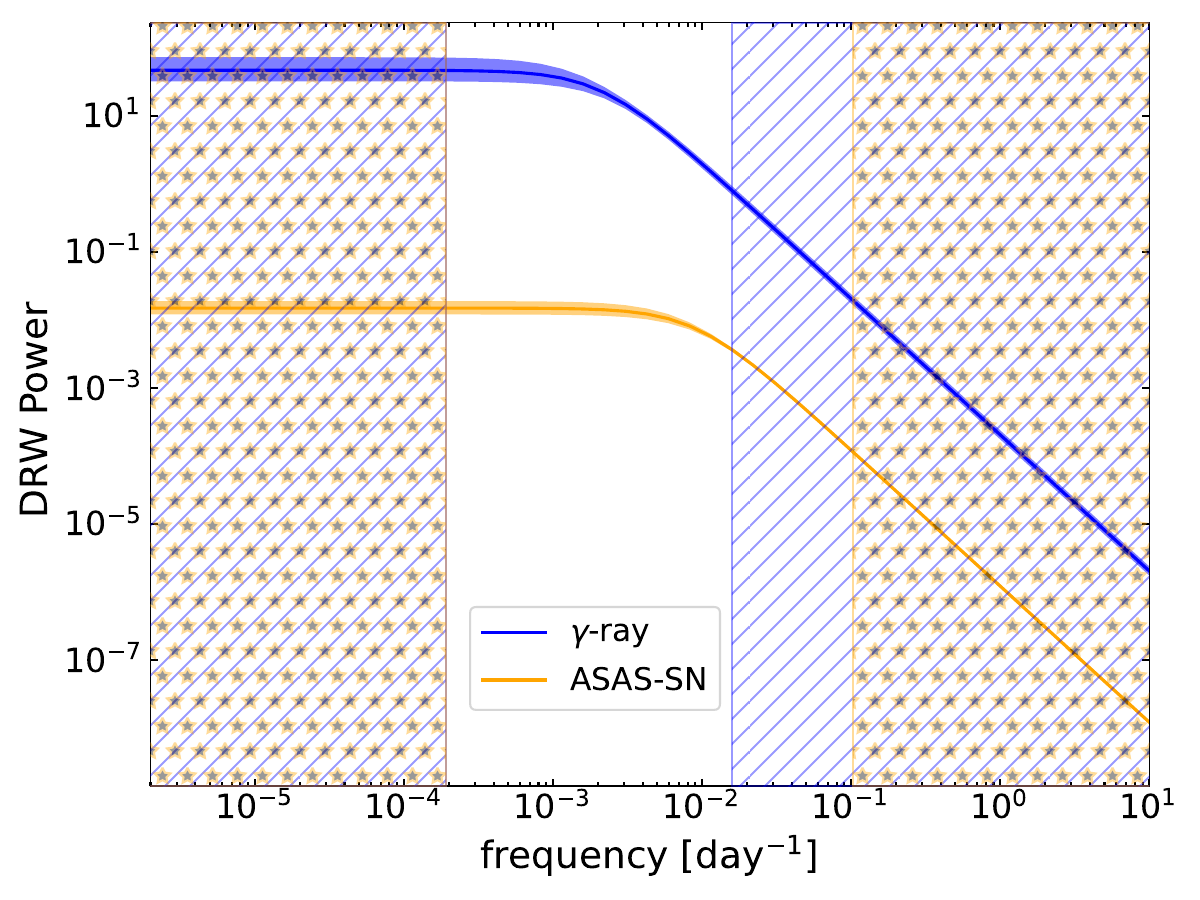} 
    \caption{This figure presents the DRW PSDs obtained from $\gamma$-ray and ASAS-SN observations, along with their 1$\sigma$ confidence intervals. The two shaded regions highlight biased regions due to observational limitations, such as finite length and the mean cadence of the light curve. The regions with orange star symbols represent the invalid areas in $\gamma$-ray PSD, while the blue hatch line regions in PSD indicate limitations imposed by the ASAS-SN light curve’s duration and mean cadence. }
    \label{Fig-DRW_PSDs}    
\end{figure}

\begin{figure*}
    \centering
    \includegraphics[width=0.45\textwidth]{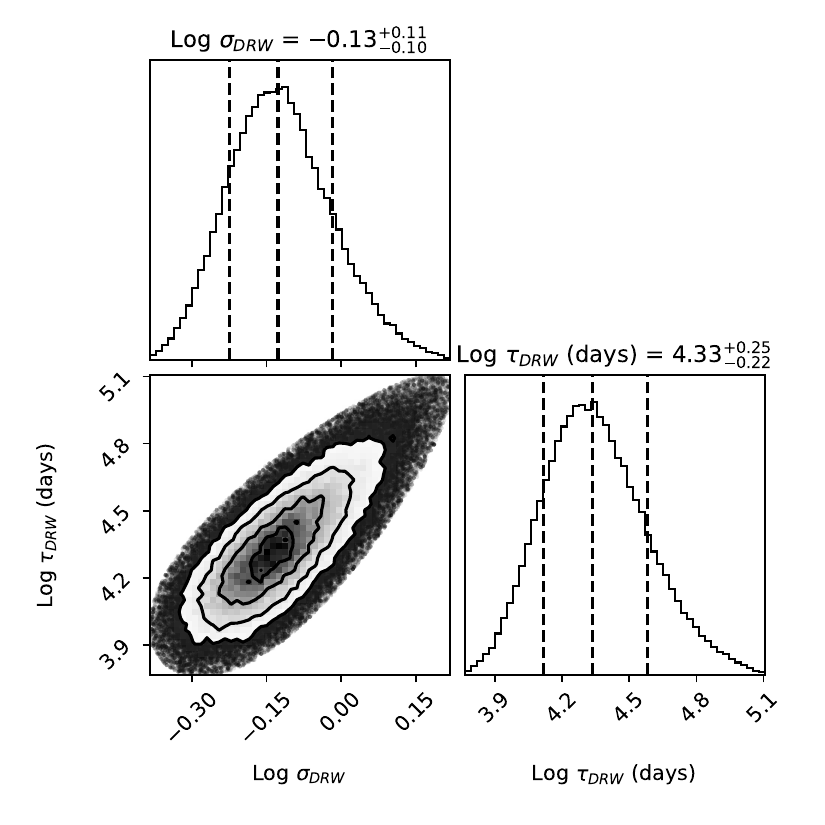} \hspace{1pt}
    \includegraphics[width=0.45\textwidth]{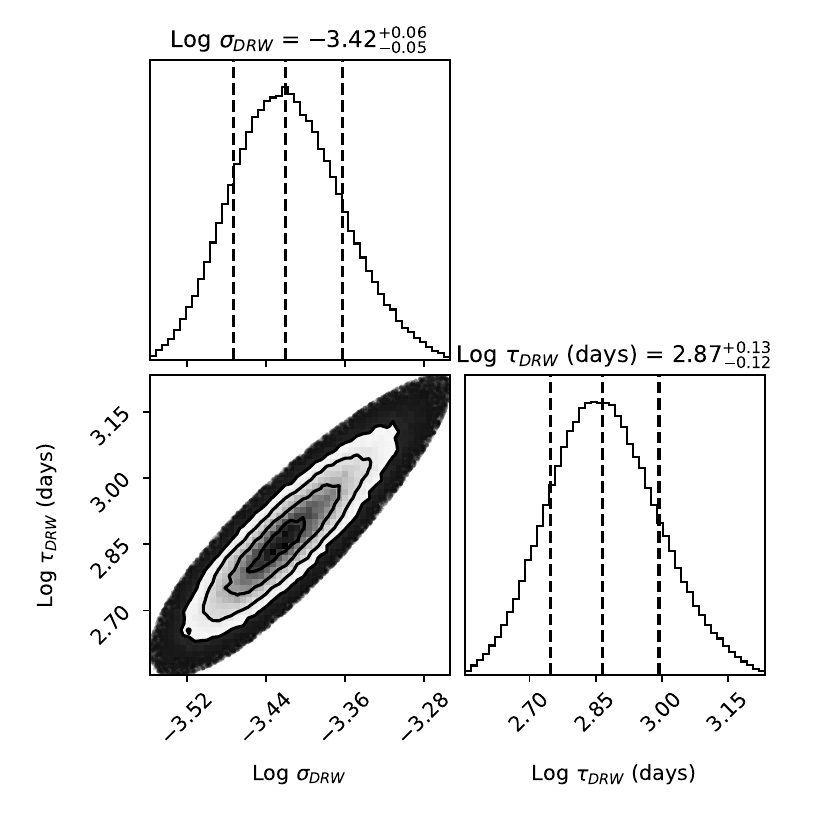}
    \caption{The figure displays the posterior probability distributions of the DRW model parameters, obtained from the $\gamma$-ray light curve (left panel) and the ASAS-SN light curve (right panel). }
    \label{Fig-corner_plots}    
\end{figure*}

\begin{figure*}
    \centering
    \includegraphics[width=0.49\textwidth]{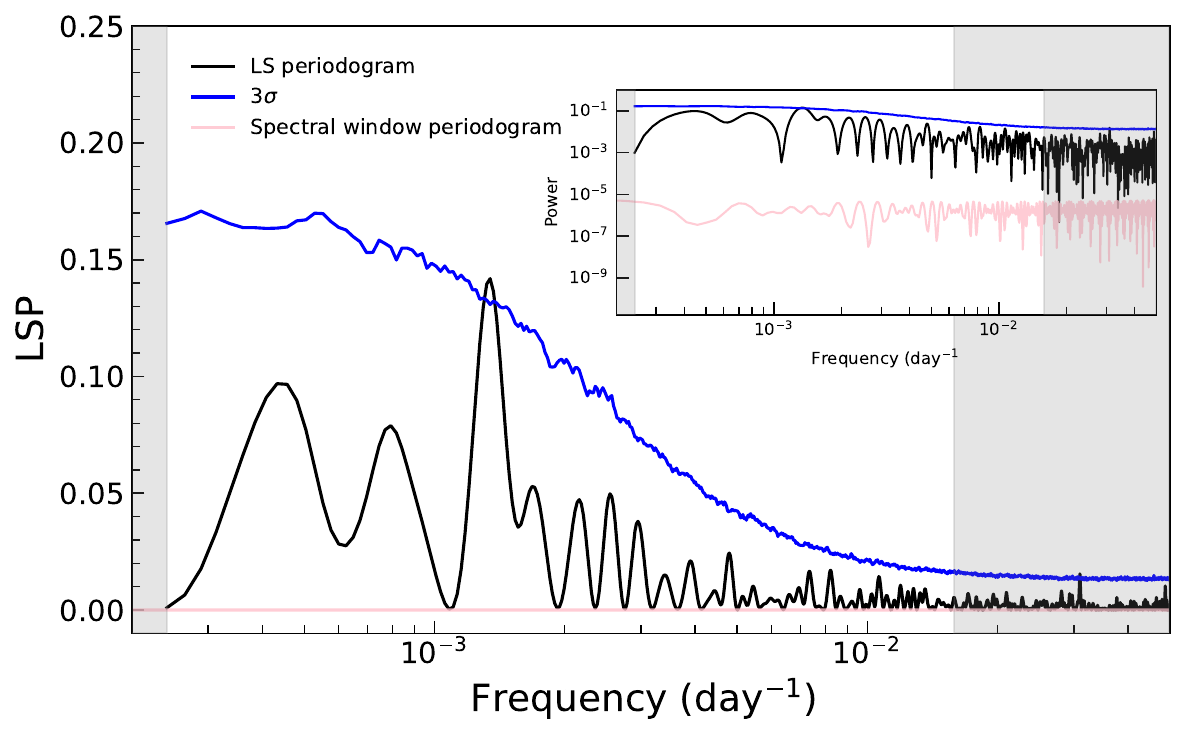} \hspace{1pt}
    \includegraphics[width=0.49\textwidth]{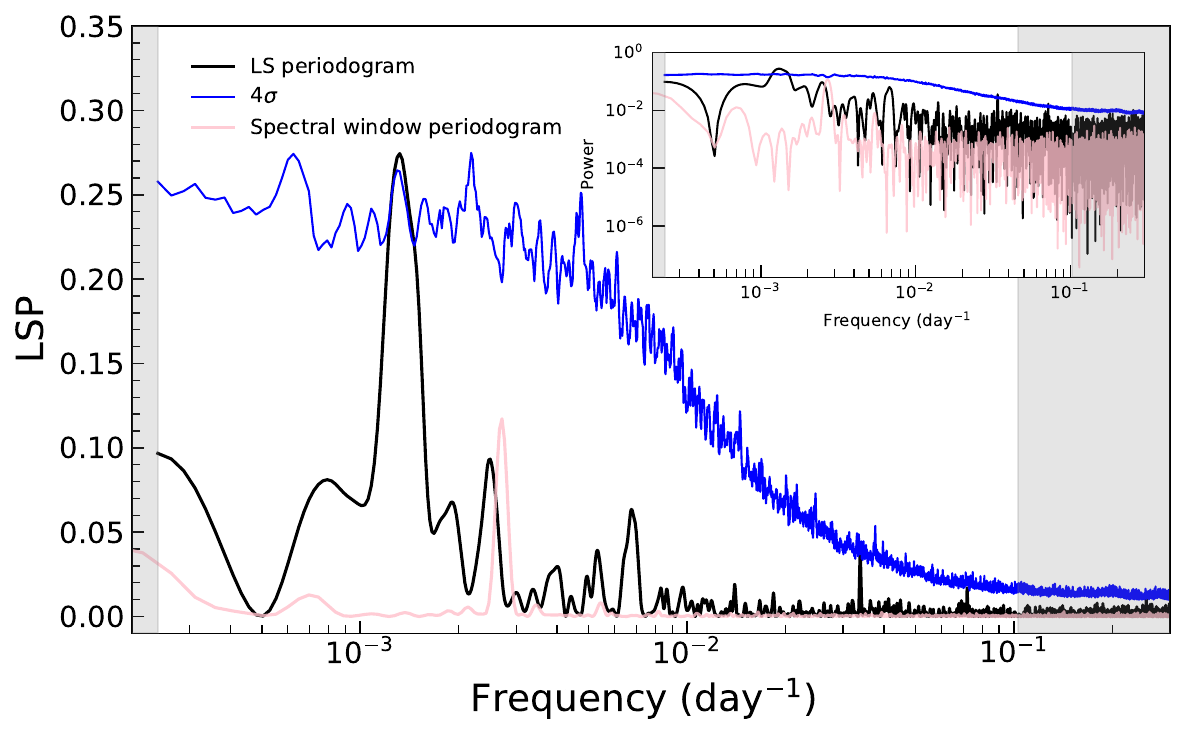}
    \caption{The Lomb-Scargle periodograms of the $\gamma$-ray (left panel) and ASAS-SN (right panel) observations of the blazar ON 246 are shown. The figure presents the LSPs of the original light curves (black) and spectral windows (pink). The significance levels (blue) of the dominant observed peaks in both periodograms are estimated using DRW synthetic light curves. The detected peaks at 0.00105 $day^{-1}$ ($\sim$950 days) and 0.00105 $day^{-1}$ ($\sim$950 days) in $\gamma$-ray and ASAS-SN observations exceed 3$\sigma$ and 4$\sigma$ significance levels, respectively. The shaded regions in both figures represent the invalid areas estimated using the mean cadence and baseline of the light curves.  Logarithmic versions of the periodograms are provided in the sub-figures.     }
    \label{Fig-DRW_LSP}    
\end{figure*}

\section{Probe the significance}\label{sec:sec5}
AGN emissions exhibit stochastic variability and are well characterized by red noise. The combination of red noise, characteristic nature, and uneven sampling in the light curve can lead to spurious peaks in the periodogram. Therefore, it is crucial to evaluate the significance of any periodic signals detected in the light curves. In the estimation of significance, we employed two different approaches.\par

The periodogram is usually represented as a power spectral density (PSD) of a form $P(\nu) \sim A \nu^{-\beta}$, where the temporal frequency is represented by $\nu$ and $\beta > 0$ represents the spectral slope. In the first approach, to calculate the statistical significance of the periodic feature, we employed the approach developed by \citet{emmanoulopoulos2013generating}. The methodology involved modeling the observed PSD using a power-law model. For that, we employed \textit{$\mathrm{DELightcurveSimulation}$}\footnote{\url{https://github.com/samconnolly/DELightcurveSimulation}} code, which involved randomizing the amplitude and phase of the Fourier components, each mimicking the characteristics of the original data, including the best-fitting power-law model slope and similar properties in terms of flux distribution. We performed a Monte Carlo simulation, generating $5\times10^4$ synthetic light curves for each case. The simulations were based on the best-fitting power-law model slopes ($\beta$ = 0.48$\pm$0.16 for $\gamma$-rays light curve and $\beta$ = 1$\pm$0.16 for ASAS-SN light curve) utilizing their flux distribution properties respectively. Each simulated light curve for both cases mimics the underlying properties of the original light curves. After that, we generated the Lomb-Scargle periodogram of each simulated light curve as we did for both original light curves. To estimate the significance level of the dominant peak in original LSPs, we calculated the 84th, 97.5th, 99.85th, and 99.995th percentiles of the 50000 simulated LSP for each frequency value, which correspond to the 1$\sigma$, 2$\sigma$, 3$\sigma$, and 4$\sigma$ significance level. In this first approach, the significance level of the dominant peak in $\gamma$-ray LSP surpasses the 3$\sigma$ and just touches the 3$\sigma$ significance level in ASAS-SN LSP. \par

In the second approach, considering that AGN variability is stochastic and well-characterized by a red noise process, we employed the Damped Random Walk (DRW) model to describe both light curves and determine the optimal model parameters, including the amplitude and damping timescale. Using the EzTao package, we simulated 10,000 synthetic light curves with a sampling rate consistent with the real observations. After generating these synthetic light curves, we computed Lomb-Scargle periodograms (LSPs) for each, following the same procedure as for the original $\gamma$-ray and ASAS-SN light curves. The significance levels were then estimated using the same methodology as described earlier. From this analysis, the peaks at 0.00134 $day^{-1}$ in $\gamma$-ray LSP, while the peak at 0.00132 $day^{-1}$ in the optical LSP surpasses the 4$\sigma$ threshold. Additionally, we calculated the spectral window periodogram by constructing a light curve with a total number of time stamps ten times larger than the original within the observed temporal frame. In this synthetic light curve, the time stamps matching the original observations were assigned a value of one, while all others were set to zero. Further, we applied the LSP method to generate the periodogram, as shown in pink in Figure \ref{Fig-DRW_LSP}.

\begin{table*}
\setlength{\extrarowheight}{8pt}
\setlength{\tabcolsep}{7pt}
\centering
\caption{Summary of the quasi-periodic signal detection using three different methodologies. Column (1) lists the 4FGL name of the blazar ON 246, while Column (2) specifies the observational waveband. Columns (3) and (4) present the QPO frequencies obtained from the Lomb-Scargle periodogram and Weighted Wavelet Z-transform methods, along with their uncertainties. The local significance of each detected QPO signal is provided in parentheses next to the corresponding frequency value. Column (5) presents the estimated local significance level of the QPO in LSP based on DRW-modeled mock light curves, and Column (6) provides the QPO frequency and significance level derived from the REDFIT analysis.}

\begin{tabular}{c c c c c c}

\toprule
\multirow{2}{*}{4FGL Name} & \multirow{2}{*}{Band} & \multicolumn{2}{c}{Monte Carlo simulation} & \multicolumn{1}{c}{Red noise modeling} & \multirow{2}{*}{REDFIT $\left(10^{-3} \ day^{-1} \right)$} \\
\cmidrule(lr){3-4} \cmidrule(lr){5-5}
 & & LSP $\left(10^{-3} \  day^{-1} \right)$ & WWZ $\left(10^{-3} \  day^{-1} \right)$ & LSP $\left(10^{-3} \  day^{-1} \right)$ & \\
(1) & (2) & (3) & (4) & (5) & (6)\\
[+2pt]
\midrule
\multirow{2}{*}{4FGL J1230.2+2517} & $\gamma$-ray &  1.34$\pm$0.1 ($>99.73\%$) & 1.32$\pm$0.15 ($>99.73\%$) & 
1.34$\pm$0.1 ($>99.73\%$) & 
1.28$\pm$0.5 ($>99\%$)\\
 & Optical &  1.32$\pm$0.2 ($>99.73\%$) & 1.31$\pm$0.2 ($>99.73\%$) & 
1.32$\pm$0.2 ($>99.99\%$) & 
1.32$\pm$0.4 ($>99\%$)\\
[+4pt]
\hline
\end{tabular}
\vspace{6pt}
\begin{minipage}{\textwidth}
\fontsize{9.5pt}{10pt}\selectfont
\textbf{Note:} The QPO peaks were fitted using a Gaussian function, and the uncertainties correspond to the half-width at half-maximum (HWHM).
\end{minipage}
\label{tab:QPO_all}
\end{table*}

\section{\textbf{Gamma-ray/optical cross correlations}}\label{sec:DCF}
We investigate the possible time lags between the $\gamma$-ray and optical light
curves of ON 246 utilizing the interpolated cross-correlation function (ICCF: \cite{peterson1998uncertainties, peterson2004central}), which is one of the commonly used methods in the time-series analysis of AGNs. As we know, AGN light curves generally are not regularly sampled in time but are discretely sampled N times at time stamps $t_i$, where $t_{i+1} - t_i = \Delta t$ for all values $1 \le i \le N-1$. ICCF method emerges as a powerful technique to estimate the time-leg between two-time series. The method uses the linear interpolation method to deal with unevenly sampled light curves and calculate the cross-correlation coefficient as a function of the time lag for two light curves:

\begin{equation}
    F_{CCF}(\tau)=\frac{1}{N} \sum_{i=1}^N \frac{\left[ L(t_i) - \bar{L} \right] \left[ C(t_i - \tau) - \bar{C} \right]}{\sigma_L \sigma_C}
\end{equation}

where N is the number of data points in the light curves L and C. Each light curve has a corresponding mean value ($\bar{L}$ and $\bar{C}$ ) and uncertainty ($\sigma_L$ and $\sigma_C$).

\begin{figure*}
    \centering
    \includegraphics[width=0.49\textwidth]{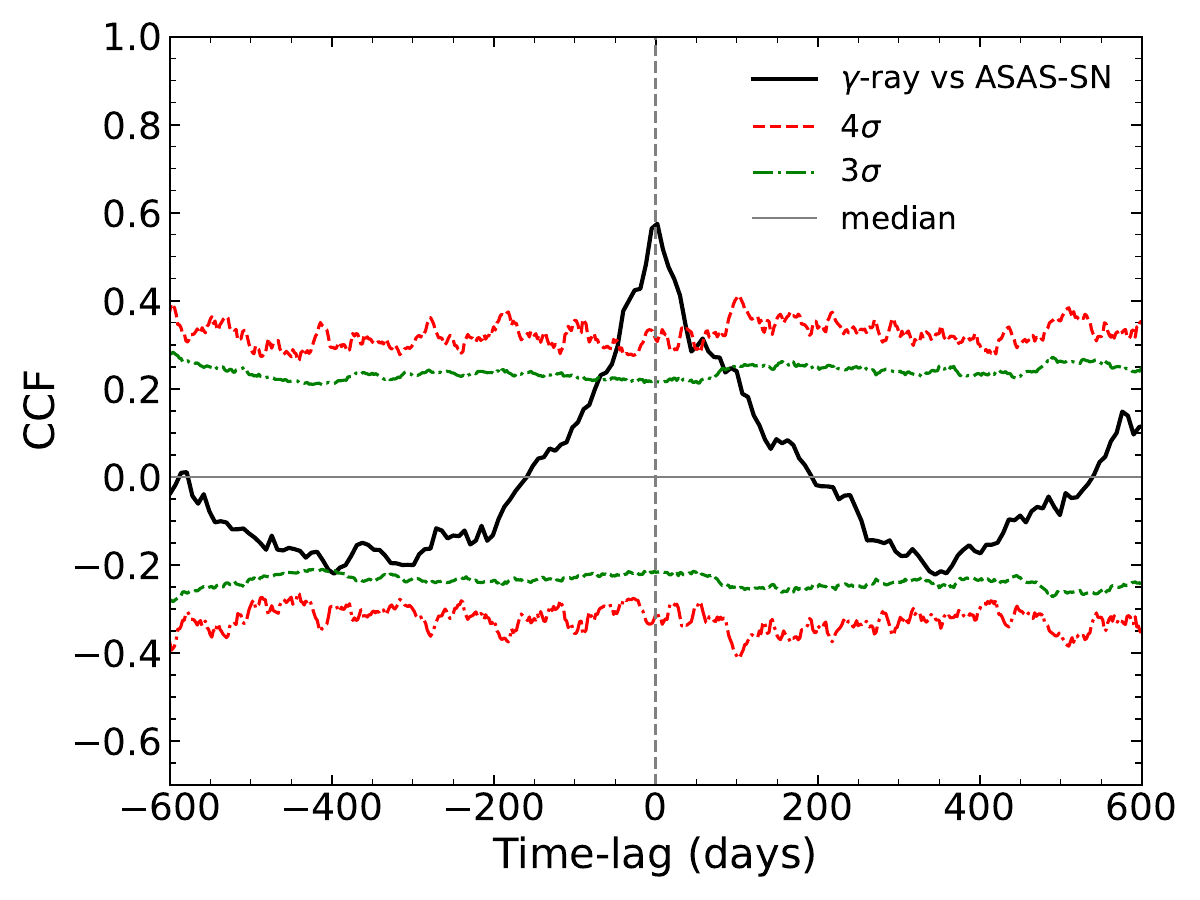}
    \hspace{1pt}
    \includegraphics[width=0.49\textwidth]{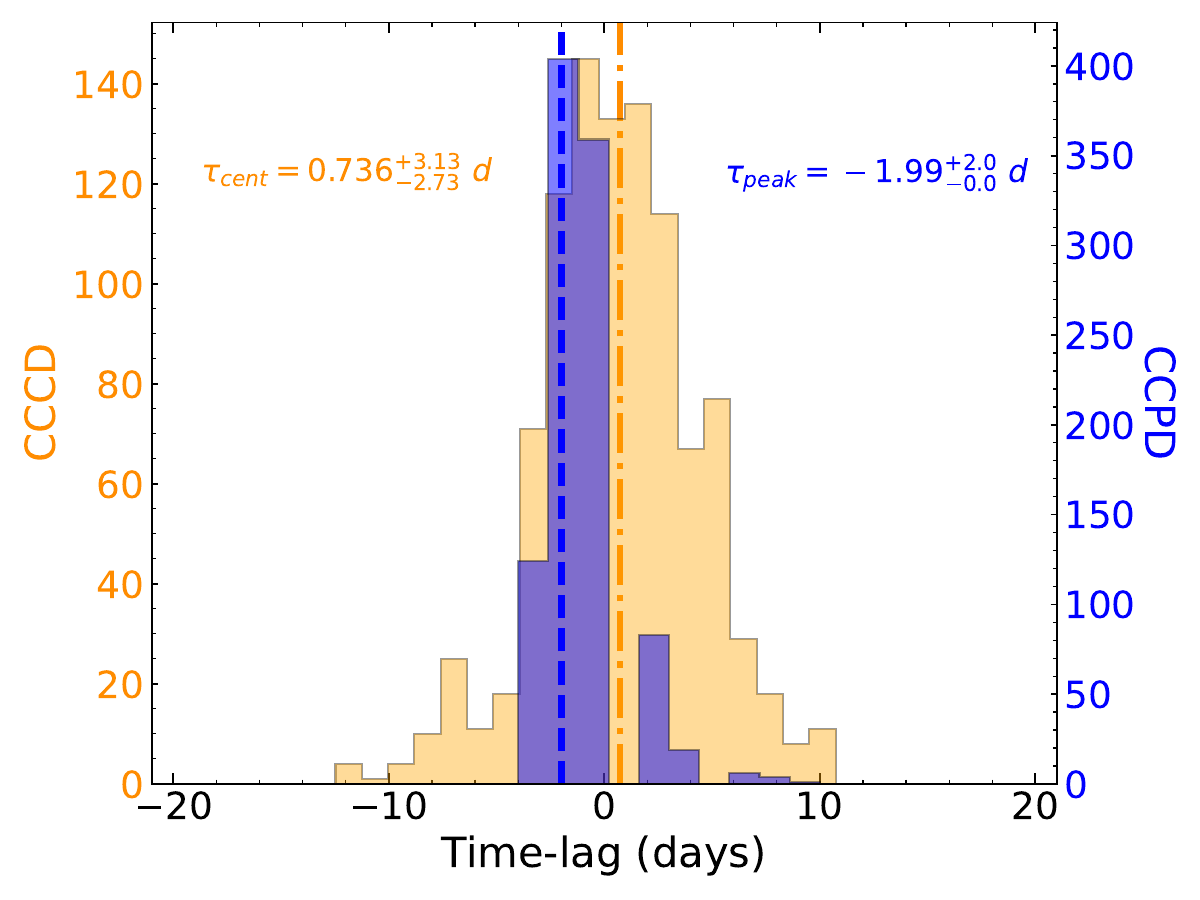}
    \caption{Cross-correlation analysis between $\gamma$-ray and optical flux variations. The left panel shows ICCF (solid black curve) with a significance level (dash red curve) of 4$\sigma$. The cross-correlation centroid distribution (CCCD) in orange and cross-correlation peak distribution (CCPD) in blue are given in the right panel, where vertical dashes in orange and blue represent the median of CCCD and CCPD, respectively.}
    \label{Fig-CCF_CCCD}    
\end{figure*}

The ICCF is evaluated for a time lag ($\tau$) in a range [-1000,1000] days with searching step $\Delta \tau$, which should be smaller than the median sampling time of the light curves. We adopted $\Delta \tau$=7 days and used the public PYTHON version of the ICCF, PYCCF \citep{2018ascl.soft05032S} in this study. As applying the ICCF to the light curves, a strong peak is obtained, and its corresponding centroid is calculated using the ICCF for the time lags around the peak. We adopted the centroid of the CCF ($\tau_{cent}$) using only time lags with r$> 0.8 r_{max}$, where $r_{max}$ is the peak value of the CCF. The 1$\sigma$ confidence on the time lag is estimated using a model-independent Monte Carlo method. We found a lag of $0.736_{-2.73}^{+3.13}$ day with the $\gamma$-ray leading the optical emissions. In addition to cross-correlation centroid distribution (CCCD), we also estimated the cross-correlation peak distribution (CCPD); see the right panel of Figure \ref{Fig-CCF_CCCD}. To access the significance, we also calculated the ICCF between the $\gamma$-ray and DRW mock optical light curves (see Section \ref{sec:sec5}); the observed findings are shown in the left panel of Figure \ref{Fig-CCF_CCCD}. After simulating the 10000 mock ICCFs between $\gamma$-ray and optical mock light curves, we calculated the significance level of the ICCF peak observed in correlation analysis between the observed $\gamma$-ray and ASAS-SN lightcurves. In the left panel of Figure \ref{Fig-CCF_CCCD}, a red dashed curve represents the 99.9999$\%$, corresponding to 4$\sigma$, significance level. The observed finding indicates a significant correlation with almost zero-day lag between $\gamma$-ray and optical emissions, suggesting a common origin of them \citep{cohen2014temporal}.

\section{Black hole mass estimation}\label{sec:BH-mass}
The mass of the central black hole in an AGN is one of the most crucial parameters for understanding its central engine. Several methods have been developed to estimate black hole mass, including stellar and gas kinematics and reverberation mapping \citep{gupta2008periodic}. The stellar and gas kinematics approach requires high spatial resolution spectroscopic observations to resolve the gravitational influence of the black hole. Reverberation mapping relies on detecting high-ionization emission lines from regions close to the black hole. 
In addition to the black hole mass estimation methods mentioned above, the variability timescales can also serve as a useful tool for determining the mass of an AGN's black hole. The timescales of high-amplitude variations may be linked to the black hole mass, offering valuable insights into the central engine of AGNs. The shortest variability timescale provides a constraint on the size of the emitting region in blazars. In relativistic jets, the emitting region is often simplified as a "blob" with a characteristic size $D$. This size can be constrained by the relation:
\begin{equation}
    D \lesssim \frac{\delta \Delta t_{\mathrm{obs}} c}{1+z}
\end{equation}
where $\delta$ is the Doppler factor of the relativistic jet, $\Delta t_{obs}$ is the observed variability timescale, $z$ is the redshift of the blazar, and c represents the speed of light.\par

Several studies \citep{hartman1996gamma, ghisellini1996origin, xie1998massive, yang2010central} have suggested that $\gamma$-ray emissions originate from a region located a few hundred Schwarzschild radii ($r_g 
\equiv GM_* /c^2$), where G and $M_*$ are the gravitational constant and black hole mass, respectively. The emission region is typically constrained to a size of $r < 200 r_g $. \citet{yang2010central} provided a relation to estimate the lower limit of the black hole mass in terms of $\gamma$-ray luminosity, redshift, and variability timescale, which is defined as 

\begin{equation}
    \frac{\mathrm{M}}{ \ \ \mathrm{M_{\odot}}} \ge 1\times10^3 \ \frac{\Delta \mathrm{ t_{obs}}}{\mathrm{1+z}} \ \left[\frac{L_{\gamma}\mathrm{(1+z)}}{6.3\times 10^{40} \Delta \mathrm{ t_{obs}}}\right]^{\frac{1}{4+\alpha_{\gamma}}}
\end{equation}

The $\gamma$-ray luminosity is defined as 

\begin{equation}
    L_{\gamma} = 4 \pi d_L^2 (1+z)^{\alpha_{\gamma}-2} . f
\end{equation}

where $d_L$ is the luminosity distance estimated following the expression given as

\begin{equation}
    d_L = \frac{c}{H_0} \int_1^{1+z} \ \frac{1}{\sqrt{\Omega_M x^3  \ + \ 1 \ - \ \Omega_M }} \ dx
\end{equation}

from the $\Lambda-\mathrm{CDM}$ model with $\Omega_M \ ; \ (1+z)^{\alpha_{\gamma} - 2}$ represents a $K$-correction factor, and hubble constant $H_0=73 \ km \ s^{-1} \ Mpc^{-1}$. \par
The $\gamma$-ray photon number per unit energy can be defined as a power-law distribution 

\begin{equation}
    \frac{dN}{dE} = N_0 E^{-\alpha_{\gamma}}
\end{equation}

where, $\alpha_{\gamma}$ represents the photon spectral index and $N_0$ is the normalisation constant. $N_0$ can be estimated by integrating eq. (16)

\begin{equation}
    N_0 = N_{(E_L - E_U)} \left( \frac{1}{E_L} - \frac{1}{E_U}\right) \ \ , \ \ \ \mathrm{if}  \ \ \alpha_{\gamma} = 2 \ ;
\end{equation}

\begin{equation}
    N_0 = N_{(E_L - E_U)} . \frac{1 - \alpha_{\gamma}}{E_U^{1 - \alpha_{\gamma}} - E_L^{1 - \alpha_{\gamma}}} \ \ , \ \ \ \mathrm{if}  \ \ \alpha_{\gamma} \neq 2
\end{equation}

where $N_{(E_L - E_U)}$ is the integral photons in units of photon $\mathrm{cm^{-2}}$ $\mathrm{s^{-1}}$ in the energy range $E_L - E_U$, $E_L$ and $E_U$ represent the lower and upper energy limits respectively. In this study, we set $E_L$=100 MeV and $E_U$=300 GeV. Thus, the total $\gamma$-ray flux can be calculated by $f = \int_{E_L}^{E_U} \ EdN$, which elaborated as 

\begin{equation}
    f = N_{(E_L - E_U)} \left( \frac{1}{E_L} - \frac{1}{E_U}\right)ln \frac{E_U}{E_L} \ \ , \ \ \ \mathrm{if}  \ \ \alpha_{\gamma} = 2 \ ;
\end{equation}

\begin{equation}
    f = N_{(E_L - E_U)} \frac{1 - \alpha_{\gamma}}{2 - \alpha_{\gamma}} \ \frac{\left( E_U^{2-\alpha_{\gamma}} \ - E_L^{2-\alpha_{\gamma}} \right)}{E_U^{1-\alpha_{\gamma}} \ - E_L^{1-\alpha_{\gamma}}} \ \ , \ \ \mathrm{otherwise}
\end{equation}
in units of GeV $\mathrm{cm^{-2}}$ s$^{-1}$. 

\citet{acharyya2023veritas} reported the spectral parameters of power-law distribution eq. (16), including $N_0 = (6.52 \pm 1.12 )\times 10^{-11} \ \mathrm{cm^{-2}} \ \mathrm{s^{-1}} \ \mathrm{MeV^{-1}}$ and $\alpha_{\gamma} = 1.79 \pm 0.14$. The estimated $\gamma$-ray luminosity is $\sim 1.18\times 10^{47} \ erg \ s^{-1}$ and the mass of the black hole to be $M_* > 1.42 \times 10^8 M_{\odot}$ using the eq.'s (14) and (13), respectively.

Additionally, \citet{liu2015constraints} proposed a model to determine the upper limit of the black hole mass based on variability studies, which is briefly discussed here. In this model, a blob in the jet-production region initially has a size of $D_0$ and expands to $D_R$ at a distance $R_{jet}$ from the central black hole. As the blob propagates outward along the jet, it expands with an average velocity $\left( \bar{v}_{exp} \right)$. Since the expansion velocity is non-relativistic ($\bar{v}_{exp} << \bar{v}_{jet}$, where $\bar{v}_{jet}$ is the relativistic jet velocity), we have the condition $D_0 \lesssim D_R$. Following Equation 12, we can express this relationship as: 
\begin{equation}
    D_0 \lesssim D_R \leqslant \frac{\delta \Delta t_{\mathrm{min}}^\mathrm{obs}}{1+z} c
\end{equation}
where $\Delta t_{min}^{obs}$ is the observed minimum timescale of variability. Relativistic jets are believed to originate from the inner regions of the accretion disk, close to the central black hole \citep{blandford1977electromagnetic, blandford1982hydromagnetic,meier2001magnetohydrodynamic}. The inner radius of the disk is typically assumed to be near the marginally stable orbit, also known as the innermost stable circular orbit (ISCO). The ISCO radius ($r_{ms}$) is expressed in terms of the gravitational radius ($r_g$) and the dimensionless spin parameter $j = J/J_{max}$, where the maximum angular momentum is given by $J_{max}=GM_*^2 / c$, as defined in Equation 5 of \citet{liu2015constraints}. For a Schwarzschild black hole ($j=0$, non-rotating), the ISCO is located at $r_{ms}=6r_g$, where $r_g=GM_* / c^2$. The corresponding size of the emitting region is $D_{ms}=12r_g$.\par
In the case of Kerr black hole ($j$=1, maximally rotating), for prograde orbits, the emitting region is $D_e = 2r_e = 4r_g$, where $r_e$ represents the equatorial boundary of the ergosphere. These estimated sizes are consistent with findings from general relativistic magnetohydrodynamic (GRMHD) simulations \citep{meier2001magnetohydrodynamic}. Thus, the size of the blob spans from $D_0=4-12 r_g$ for $0\leqslant j \lesssim 1$. Following equation 13, we have (as given in equations 6a and 6b in \cite{liu2015constraints} ) 
\begin{align}
    M_{\mathrm{var}}^{\mathrm{Sch}} &\lesssim 1.695 \times 10^4 \frac{\delta \Delta t_{\mathrm{min}}^\mathrm{obs}}{1+z} M_{\odot} \quad , \quad (j \sim 0) \\[10pt]
    M_{\mathrm{var}}^{\mathrm{Kerr}} &\lesssim 5.086 \times 10^4 \frac{\delta \Delta t_{\mathrm{min}}^\mathrm{obs}}{1+z} M_{\odot} \quad , \quad (j \sim 1)
\end{align}
where $\Delta t_\mathrm{min}^\mathrm{obs}$ is in units of seconds.\par
\citet{acharyya2023veritas} reported the fasted observed variability timescale in $\gamma$-rays of 6.2$\pm$0.9 hr around MJD 57178. We utilized this timescale to constrain the black hole mass. The calculated black hole masses are 
$M_\mathrm{var}^\mathrm{Sch} \lesssim2.74 \times 10^9 \ M_{\odot}$ and $M_\mathrm{var}^\mathrm{Kerr} \lesssim8.22\times 10^9 \ M_{\odot}$
. In the calculation, we adopted the Doppler factor value, $\delta$=9.6 \citep{zhou2021intrinsic, acharyya2023veritas}. The uncertainties in the estimated black hole masses were determined based on the errors in the variability timescale. However, the lack of precise information regarding the exact location of the emission region relative to the central black hole could introduce additional uncertainties, potentially increasing the error bars on the mass estimates. \par

By combining the estimates of the lower and upper limits of the black hole mass, we obtain a mass range of $(0.142 < M_* < 8.22)\times 10^9 \ M_{\odot} $. Notably, the lower limit derived from the variability study is higher than the estimate reported by \citet{pei2022estimation}.\par

In recent years, several studies have demonstrated a correlation between variability timescales and black hole mass. \citet{burke2021characteristic} identified a relationship between optical timescales in accretion disks and black hole masses, spanning the entire mass range of supermassive black holes and even extending to stellar-mass BH systems. This correlation has been further explored across different electromagnetic bands, including $\gamma$-rays \citep{ryan2019characteristic, zhang2022characterizing, zhang2023gaussian, sharma2024probing, zhang2024discovering, sharma2024microquasars}, X-rays \citep{zhang2024discovering}, and sub-millimeter wavelengths \citep{chen2023testing}. Notably, the observed $\gamma$-ray variability timescales in AGNs have been found to overlap with those in the optical band \citep{burke2021characteristic}, suggesting a link between non-thermal jet emissions and thermal emissions from accretion disks. Thus, by utilizing the relationship between the rest-frame variability timescale $\left(\tau_{rest}^{Damping}\right)$ and BH mass $\left(M_{BH}\right)$ as established by \citet{burke2021characteristic}, $\tau_{rest}^{damping}=107_{-12}^{+12} \left(\frac{M_{BH}}{10^8 M_{\odot}}\right)^{0.38_{-0.04}^{+0.05}} \ \mathrm{days}$, the mass of the SMBH can be estimated.

In our study, we estimated the mass of the SMBH to be approximately $(7.3 \pm 6) \times 10^9 M_{\odot}$ based on a rest-frame damping timescale of $\sim$547 days observed in $\gamma$-ray emission of the blazar ON 246. While the uncertainty in the SMBH mass estimation is considerably high, the derived mass value is less and close to the upper limit of $M_{var}^{Kerr}$ and higher than the $M_{var}^{Sch}$, as calculated using the model proposed by \citet{liu2015constraints}. Further, the SMBH mass of ON 246, estimated using the optical timescale derived from DRW modeling, is $(1.56 \pm 0.66)\times 10^8 M_{\odot}$. \par
The estimated black hole mass of ON 246, derived from variability characteristics in both energy bands, falls within the range of $(0.142 - 8.22) \times 10^9 \ M_{\odot}$. The estimated mass also lies in the range (10$^8$--10$^9$ M$_{\odot}$) of mass of SMBH for FSRQ as derived in various studies \citep{2008MNRAS.387.1669G, 2013A&A...560A..28C, 10.1093/mnras/stu755, Paliya_2021, 10.1093/mnras/stae765} using various sample of FSRQs. 
Using the gamma-ray variability time \citet{pei2022estimation} have also estimated the mass of the black hole, but the value is close to $\sim$8.08$\times$10$^{7}$ M$_{\odot},$ which is much smaller than the value estimated above. This discrepancy is because in \citet{pei2022estimation}, authors have used a fixed variability time of 1 day and the Doppler factor ($\delta$) as a very small value of 0.48.

\section{RESULT AND DISCUSSION}\label{sec:sec6}
In this study, we investigated quasi-periodic signals in the $\gamma$-ray and optical emissions of the blazar ON 246 (4FGL J1230.2+2517) over the period of 11.6 yr, from MJD 55932 to 60081, employing three different methodologies as outlined in Section \ref{sec:sec3}. Our analysis reveals a distinct quasi-periodic signal in the $\gamma$-ray emissions of the blazar ON 246, with periods of approximately 746 days, 757 days, and 781 days, as identified using the Lomb-Scargle Periodogram (LSP), Wavelet Weighted Z-transform (WWZ), and REDFIT analysis, respectively. The significance of the detected periodicity was assessed through two independent approaches: Monte Carlo simulations and stochastic modeling using the damped random walk (DRW) model. The dominant peaks detected in both LSP and WWZ have a local significance level above 3$\sigma$, based on Monte Carlo simulations, and exceed 3$\sigma$ when evaluated using DRW modeling. The uncertainties in the observed periods were estimated as the half-width at half-maximum (HWHM) by fitting the peak profiles with a Gaussian function. The QPO detected in $\gamma$-ray emissions is further corroborated by an independent analysis of the optical light curve. Using the same methodologies, we searched for oscillatory signals in the optical data and assessed their significance. A similar periodicity of $\sim$757, $\sim$763, and $\sim$757 days were found in LSP, WWZ, and REDFIT analyses, respectively, with a significance level exceeding 3$\sigma$, reinforcing the presence of the detected QPO.\par

In addition, we also employed the interpolation cross-correlation function (ICCF) to analyze the correlation between $\gamma$-ray and optical emissions. As display in the figure \ref{Fig-CCF_CCCD}, it can be seen that the correlation coefficient is maximum at near zero time-lag, with $\tau_{cent}=0.736_{-2.73}^{+3.13}$ d, have a significance level exceeding 4$\sigma$, indicating that the variability features between these two bands are similar in nature and is believed to be originated by the lepton single-zone scenario of blazar emission \citep{cohen2014temporal}. Studying correlations across multiple wavebands is crucial for gaining deeper insights into the emission mechanisms of these systems. A strong correlation between low-energy and high-energy emissions can be well explained by the leptonic model. In this scenario, low-energy radiation originates from synchrotron emission, while the same seed photons undergo synchrotron self-Compton (SSC) and inverse Compton scattering to produce high-energy radiation, leading to a significant correlation between the two energy bands \citep{maraschi1992jet, giommi1999synchrotron, tagliaferri2003wide, zheng2013modelling, liao2014multiwavelength, li2016multiband}.\par
Conversely, when high-energy emission is generated through the external Compton process, which involves seed photons originating outside the jet \citep{malmrose2011emission, liao2014multiwavelength}, the correlation between low-energy and high-energy emissions tends to weaken or become insignificant. Our findings reveal a strong correlation between the optical and $\gamma$-ray emissions of the blazar ON 246, supporting the leptonic model's predictions. \citet{acharyya2023veritas} observed a strong correlation between MeV and GeV emissions with a peak at zero time-lag, suggesting the same origin, and also found the positive time-lags with radio and optical emissions.\par

In this study, we employed three different methods to estimate the black hole (BH) mass of the blazar ON 246 based on its variability characteristics. First, we determined the lower and upper limits of the BH mass using the models proposed by \citet{yang2010central} and \citet{liu2015constraints}, respectively. The minimum variability timescale observed in the $\gamma$-ray band yielded a lower mass limit of $M_* > 0.142 \times 10^9 \ M_{\odot}$. For the upper limit, we obtained $M_* < 2.74 \times 10^9 \ M_{\odot}$ for a Schwarzschild black hole, and $M_* < 8.22 \times 10^9 \ M_{\odot}$ for a maximally rotating Kerr black hole. In a third approach, we utilized the damping timescales of $\gamma$-ray and optical light curves, applying an empirical correlation between the rest-frame damping timescale ($\tau_{\rm rest}^{\rm Damping}$) and the BH mass, as established by \citet{burke2021characteristic}. This method yielded BH mass estimates of $(7.3 \pm 6) \times 10^9 \ M_{\odot}$ from $\gamma$-ray variability and $(1.56 \pm 0.66) \times 10^8 \ M_{\odot}$ from optical variability. Combining these results, we constrain the BH mass of ON 246 to lie within the range of approximately $(0.142 - 8.22) \times 10^9 \ M_{\odot}$.


\subsection{Potential mechanisms for QPO}
Several physical models have been proposed in the literature to explain the phenomenon of quasi-periodic oscillations (QPOs) in blazars. These include scenarios involving supermassive binary black hole (SMBBH) systems, precession or helical motion of relativistic jets (geometric effects), and instabilities in the accretion flow within the disk. A more detailed discussion on each of these scenarios is given below.\par
The supermassive binary black hole (SMBBH) scenario provides a plausible explanation for long-term flux modulations observed in blazars \citep{sillanpaa1988oj, xie2008periodicity, valtonen2008massive, villforth2010variability, graham2015possible, wang2017nearly, gong2022quasiperiodic, gong2024detection}. Several sources have been identified as potential candidates exhibiting long-term periodic flux modulations across multiple wavebands. Notable examples include a $\sim$12-year QPO in OJ 287 \citep{sillanpaa1988oj, valtonen2008massive, villforth2010variability, sandrinelli2016quasi}, a $2.18 \pm 0.08$-year periodicity in PG 1335+113 \citep{ackermann2015multiwavelength}, a $1.84 \pm 0.1$-year period in PKS 1510-089 \citep{xie2008periodicity}, and a 3-year periodicity in 3C 66A \citep{otero2020quasi}. Several other sources are also considered potential candidates exhibiting long-term periodic flux variations consistent with the SMBBH scenario. These periodic modulations in blazars are often attributed to the orbital motion of the secondary black hole within the binary system.\par 

\citet{rieger2004geometrical} investigated the potential geometrical origins of periodicity in blazar-type sources, proposing that periodic variations in emission can result from three possible mechanisms, including orbital motion in a binary black hole system, jet precession, and intrinsic jet rotation. The precessional-driven ballistic motion is unlikely to produce observable periods shorter than several decades, While the orbital motion in a close SMBBH system produces a period of $P_{obs}\gtrsim$ 10 days and Newtonian-driven precession in a close SMBBH can be a possible mechanisms of periodicity $P_{obs} \gtrsim 1$ yr. Therefore, the observed periodic flux modulations in emission from these sources canbe  reasonably explained by orbital-driven (nonballistic) helical motion in a close SMBBH system.\par
In our study, the observed oscillation period is $P_{obs}$=746 days, which is significantly shorter than the actual physical period $P_d$ due to the light-travel time effect \citep{rieger2004geometrical}. The period is related by the equation, $P_d = \frac{P_{obs} \Gamma^2}{1+z}$, where $\Gamma$ is the bulk Lorentz factor. To estimate $\Gamma$, we used a relation from \citet{1995PASP..107..803U, li2024optical}, $\Gamma \le \frac{1}{2} \left(\delta + \frac{1}{\delta}\right)$, which is basically gives the lower limit to the $\Gamma$ at a given value of $\delta$. This condition is valid only for $\delta >1$. Using a Doppler factor of $\delta$=9.6 \citep{zhou2021intrinsic}, we estimate the lower limit of the Lorentz factor as $\Gamma \geq 4.85$. Consequently, the intrinsic period of the blazar ON 246 is $\sim$36.28 years. Based on this corrected QPO period, the estimated mass of the primary black hole is 
\begin{equation}
    M \simeq P_{\mathrm{corrected, yr}}^{8/5} \ R^{3/5} \ 10^6 \ M_{\odot}
\end{equation}
where $P_{\mathrm{corrected}}$ is the real physical period in units of years, R represents the mass ratio of the primary black hole to the secondary black hole, $R=\frac{M}{m}$. Previous studies \citep{2013A&A...557A..85R, 2021PASP..133b4101Y, Gong_2022} have suggested that the $R$ lies within the ranges of 4–10.5, 1–100, and 10–100, respectively. In our analysis, we adopt $R$ within the range of 1–100 to calculate the mass of the primary black hole. Based on this, the estimated mass is found to be in the range of $ M \sim 3.12\times 10^8 - 5 \times 10^9 M_{\odot} $. The estimated BH mass range is comparable with the derived mass range in sec. \ref{sec:BH-mass}, indicating that the observed year-like QPO is likely caused by non-ballistic helical motion driven by the orbital dynamics of a close SMBBH system.\par
Additionally, instabilities in the accretion flow within the disk may also contribute to the flux modulations in blazars. In this scenario, oscillations in the innermost regions of the accretion disk or Kelvin-Helmholtz instabilities could lead to quasi-periodic plasma injections into the jet, as a result producing quasi-periodic oscillations in the jet emissions \citep{gupta2008periodic, 10.1093/mnras/stu1135, bhatta2016detection, sandrinelli2016quasi, tavani2018blazar}. The mass of the SMBH can be estimated using the following equation
\begin{equation}
    M=\frac{3.23 \ \times \ 10^4 \ \delta \ P_{obs} }{\left(r^{3/2}+a\right) \left(1+z\right)} \ M_{\odot}
\end{equation}
where $P_{obs}$ is in units of seconds, $\delta$ is the Doppler factor, parameter $r$ is the radius of this source zone in untis of $GM/c^2$, parameter $a$ is the spin parameter of SMBH, and $z$ is redshift. In our study, using equation 17, the estimated mass of the SMBH is $1.02 \times 10^{12} M_{\odot}$ for a Schwarzschild black hole with $r$=6 and $a$=0, while for a maximal Kerr black hole with $r$=1.2 and $a$=0.9982, the estimated mass of SMBH is $6.52 \times 10^{12} M_{\odot}$. The estimated masses of SMBH significantly exceed the black hole mass calculated in Section \ref{sec:BH-mass}. Therefore, the observed flux modulations in both bands are unlikely to be associated with this scenario.\par

As previously discussed, blazar emission is primarily jet-dominated, and variability in jet emission can also arise due to geometric effects. One such scenario involves a plasma blob moving along a helical trajectory within the jet, causing observed emission variations \citep{villata1999helical, rieger2004geometrical, Li_2009, Li_2015, mohan2015kinematics, 2017MNRAS.465..161S, zhou201834, otero2020quasi, gong2022quasiperiodic, gong2023two, prince2023quasi, sharma2024detection}. In this geometrical model, emission is enhanced due to relativistic beaming, and as the blob follows a helical path within the jet, the viewing angle changes over time. Such helical jet trajectories can result from jet bending, wiggling motion, or the presence of helical magnetic fields within the jet, leading to periodic emission patterns. Consequently, the blob’s motion causes a continuous variation in the viewing angle relative to the observer’s line of sight, which can be characterized as
\begin{equation}
    \mathrm{cos} \ \theta_{\mathrm{obs}}(t)= \mathrm{sin} \phi \ \mathrm{sin} \psi \ \mathrm{cos} (2\pi t / P_{\mathrm{obs}}) \ + \ \mathrm{cos} \phi \ \mathrm{cos} \psi
\end{equation}
where $\phi$ represents the pitch angle between the blob motion and the jet axis, $\psi$ is the viewing angle or inclination angle between the observer's line of sight and jet axis, and $P_{\mathrm{obs}}$ represents the observed period of oscillations in the light curve. Due to the helical motion of emitting region in the jet, the Doppler factor undergoes temporal variations, defined as 
\begin{equation}
    \delta = \frac{1}{\Gamma \left(1 \ -\ \beta \ \mathrm{cos} \theta_{\mathrm{obs}}(t)\right)}
\end{equation}
where, $\Gamma = \frac{1}{\sqrt{1 - \beta^2}}$ represents the bulk Lorentz factor and $\beta = v_{\mathrm{jet}/c}$. Following this, the observed flux is defined as $F_{\nu} \propto \delta^3$. The relationship between the observed period and rest frame period of blob can be defined by the following expression:
\begin{equation}
    P_{\mathrm{rest}}=\frac{P_{\mathrm{obs}}}{1 \ - \ \beta \ \mathrm{cos} \psi \ \mathrm{cos} \phi}
\end{equation}
The blazer ON 246 is BL Lac type, therefore, we considered typical values of $\phi=2^{\circ}, \ \psi = 5^{\circ}$ \citep{2017MNRAS.465..161S, zhou201834, sarkar2019long, prince2023quasi}, and $\Gamma=4.85$ as estimated above. The period of oscillations in the rest frame to be $P_{\mathrm{rest}} \sim$ 78.85 yr.\par

As discussed earlier, \citet{rieger2004geometrical} outlined three potential mechanisms, one of which suggests that if the observed periodicity exceeds 1 year, it can be reasonably attributed to the helical motion of the jet driven by the orbital motion of an SMBBH system. Using the expression in Equation 16 and periodicity in rest frame, we estimate the mass of the primary black hole to be $M \sim 1.083 \times 10^9, \ 4.314 \times 10^9, \ 1.71 \times 10^{10} M_{\odot}$ for R=1,10, and 100, respectively. The estimated primary black hole mass in SMBBH for R=1,10 is consistent with the derived mass value utilizing different approaches as described in section \ref{sec:BH-mass}. Therefore, our comprehensive analysis suggests that the observed periodicity is most likely attributed to the jet being driven by the orbital motion of the SMBBH system.

\section{CONCLUSION}\label{sec:sec7}
In this study, we investigated the $\gamma$-ray and optical emission of the blazar ON 246 over the period of 11.6 yr (MJD 55932–60081). The key findings of our analysis are summarized as follows:
\begin{itemize}
    \item We identified a potential quasi-periodic signal of approximately 746 days in the $\gamma$-ray and optical emissions of blazar ON 246 with a significance level exceeding 3$\sigma$.
    \item Cross-correlation analysis reveals a strong correlation between $\gamma$-ray and optical emissions, indicating a common origin for both. Additionally, we assessed the significance of the peak in the cross-correlation plot, finding it to be at the 4$\sigma$ level. 

    \item We estimated the black hole mass range of ON 246 to be $(0.142 - 8.22) \times 10^9 \ M_{\odot}$ based on the shortest variability timescale observed in the $\gamma$-ray band. Furthermore, using the rest-frame damping timescales in the $\gamma$-ray and optical emissions, the black hole masses were found to be $(7.3 \pm 6) \times 10^9 \ M_{\odot}$ and $(1.56 \pm 0.66) \times 10^8 \ M_{\odot}$, respectively. Both estimated values fall within the derived black hole mass range.

    
    \item To explain the observed QPO, we explored various scenarios that could potentially account for the flux modulation with a period of $\sim$746 days in both light curves. Our analysis suggests that a non-ballistic jet driven by the orbital motion of a close supermassive binary black hole (SMBBH) system is a plausible explanation for the long-term periodic variations observed in the light curves. 
\end{itemize}

\paragraph{\textbf{Acknowledgments}}
A. Sharma is grateful to Prof. Sakuntala Chatterjee at S.N. Bose National Centre for Basic Sciences for providing the necessary support to conduct this research. RP acknowledges the support of the IoE seed grant from BHU.



\paragraph{\textbf{Data Availability Statement}}
This research utilizes publicly available data of ON 246 obtained from the Fermi-LAT data server provided by NASA Goddard Space Flight Center (GSFC): \url{https://fermi.gsfc.nasa.gov/ssc/data/access/} and ASAS-SN observations from \url{https://asas-sn.osu.edu}. 
\bibliography{sample7}{}

\begin{thebibliography}{}
\expandafter\ifx\csname natexlab\endcsname\relax\def\natexlab#1{#1}\fi
\providecommand{\url}[1]{\href{#1}{#1}}
\providecommand{\dodoi}[1]{doi:~\href{http://doi.org/#1}{\nolinkurl{#1}}}
\providecommand{\doeprint}[1]{\href{http://ascl.net/#1}{\nolinkurl{http://ascl.net/#1}}}
\providecommand{\doarXiv}[1]{\href{https://arxiv.org/abs/#1}{\nolinkurl{https://arxiv.org/abs/#1}}}

\bibitem[{A. Acharyya {et~al.}(2023)Acharyya, Adams, Archer, Bangale, Benbow, Brill, Christiansen, Chromey, Errando, Falcone, {et~al.}}]{acharyya2023veritas}
Acharyya, A., Adams, C., Archer, A., {et~al.} 2023, \bibinfo{title}{VERITAS discovery of very high energy gamma-ray emission from S3 1227+ 25 and multiwavelength observations,} The Astrophysical Journal, 950, 152

\bibitem[{M. Ackermann {et~al.}(2015)Ackermann, Ajello, Albert, Atwood, Baldini, Ballet, Barbiellini, Bastieri, Gonzalez, Bellazzini, {et~al.}}]{ackermann2015multiwavelength}
Ackermann, M., Ajello, M., Albert, A., {et~al.} 2015, \bibinfo{title}{Multiwavelength evidence for quasi-periodic modulation in the gamma-ray blazar PG 1553+ 113,} The Astrophysical Journal Letters, 813, L41

\bibitem[{S. Adhikari {et~al.}(2024)Adhikari, Penil, Westernacher-Schneider, Dominguez, Ajello, Buson, Rico, \& Zrake}]{adhikari2024constraining}
Adhikari, S., Penil, P., Westernacher-Schneider, J.~R., {et~al.} 2024, \bibinfo{title}{Constraining the PG 1553+ 113 Binary Hypothesis: Interpreting Hints of a New, 22 yr Period,} The Astrophysical Journal, 965, 124

\bibitem[{J. Aleksi{\'c} {et~al.}(2011)Aleksi{\'c}, Antonelli, Antoranz, Backes, Barrio, Bastieri, Gonz{\'a}lez, Bednarek, Berdyugin, Berger, {et~al.}}]{aleksic2011magic}
Aleksi{\'c}, J., Antonelli, L., Antoranz, P., {et~al.} 2011, \bibinfo{title}{MAGIC discovery of very high energy emission from the FSRQ PKS 1222+ 21,} The Astrophysical Journal Letters, 730, L8

\bibitem[{W. Alston {et~al.}(2015)Alston, Parker, Markevi{\v{c}}i{\=u}t{\.e}, Fabian, Middleton, Lohfink, Kara, \& Pinto}]{alston2015discovery}
Alston, W., Parker, M., Markevi{\v{c}}i{\=u}t{\.e}, J., {et~al.} 2015, \bibinfo{title}{Discovery of an~ 2-h high-frequency X-ray QPO and iron K$\alpha$ reverberation in the active galaxy MS 2254.9- 3712,} Monthly Notices of the Royal Astronomical Society, 449, 467

\bibitem[{W. Atwood {et~al.}(2009)Atwood, Abdo, Ackermann, Althouse, Anderson, Axelsson, Baldini, Ballet, Band, Barbiellini, {et~al.}}]{atwood2009large}
Atwood, W., Abdo, A.~A., Ackermann, M., {et~al.} 2009, \bibinfo{title}{The large area telescope on the Fermi gamma-ray space telescope mission,} The Astrophysical Journal, 697, 1071

\bibitem[{M.~E. Aydin(2017)Aydin}]{m_emre_aydin_2017_375648}
Aydin, M.~E. 2017, \bibinfo{title}{eaydin/WWZ: First release,}, v1.0.0 Zenodo, \dodoi{10.5281/zenodo.375648}

\bibitem[{N. Bade {et~al.}(1994)Bade, Fink, \& Engels}]{bade1994new}
Bade, N., Fink, H., \& Engels, D. 1994, \bibinfo{title}{New x-ray bright bl lacertae objects from the rosat all-sky survey,} Astronomy and Astrophysics (ISSN 0004-6361), vol. 286, no. 2, p. 381-388, 286, 381

\bibitem[{A. Banerjee {et~al.}(2023)Banerjee, Sharma, Mandal, Das, Bhatta, \& Bose}]{banerjee2023detection}
Banerjee, A., Sharma, A., Mandal, A., {et~al.} 2023, \bibinfo{title}{Detection of periodicity in the gamma-ray light curve of the BL Lac 4FGL J2202. 7+ 4216,} Monthly Notices of the Royal Astronomical Society: Letters, 523, L52

\bibitem[{M.~C. Begelman {et~al.}(1980)Begelman, Blandford, \& Rees}]{begelman1980massive}
Begelman, M.~C., Blandford, R.~D., \& Rees, M.~J. 1980, \bibinfo{title}{Massive black hole binaries in active galactic nuclei,} Nature, 287, 307

\bibitem[{F.~A. Benkhali {et~al.}(2020)Benkhali, Hofmann, Rieger, \& Chakraborty}]{benkhali2020evaluating}
Benkhali, F.~A., Hofmann, W., Rieger, F., \& Chakraborty, N. 2020, \bibinfo{title}{Evaluating quasi-periodic variations in the $\gamma$-ray light curves of Fermi-LAT blazars,} Astronomy \& Astrophysics, 634, A120

\bibitem[{G. Bhatta(2021)Bhatta}]{bhatta2021characterizing}
Bhatta, G. 2021, \bibinfo{title}{Characterizing long-term optical variability properties of $\gamma$-ray-bright blazars,} The Astrophysical Journal, 923, 7

\bibitem[{G. Bhatta \& N. Dhital(2020)Bhatta \& Dhital}]{bhatta2020nature}
Bhatta, G., \& Dhital, N. 2020, \bibinfo{title}{The nature of $\gamma$-ray variability in blazars,} The Astrophysical Journal, 891, 120

\bibitem[{G. Bhatta {et~al.}(2016)Bhatta, Zola, Ostrowski, Winiarski, Og{\l}oza, Dr{\'o}{\.z}d{\.z}, Siwak, Liakos, Kozie{\l}-Wierzbowska, Gazeas, {et~al.}}]{bhatta2016detection}
Bhatta, G., Zola, S., Ostrowski, M., {et~al.} 2016, \bibinfo{title}{Detection of possible quasi-periodic oscillations in the long-term optical light curve of the BL Lac object OJ 287,} The Astrophysical Journal, 832, 47

\bibitem[{R. Blandford \& C.~F. McKee(1982)Blandford \& McKee}]{blandford1982reverberation}
Blandford, R., \& McKee, C.~F. 1982, \bibinfo{title}{Reverberation mapping of the emission line regions of Seyfert galaxies and quasars,} Astrophysical Journal, Part 1, vol. 255, Apr. 15, 1982, p. 419-439. Research supported by the Alfred P. Sloan Foundation, 255, 419

\bibitem[{R. Blandford {et~al.}(2019)Blandford, Meier, \& Readhead}]{blandford2019relativistic}
Blandford, R., Meier, D., \& Readhead, A. 2019, \bibinfo{title}{Relativistic jets from active galactic nuclei,} Annual Review of Astronomy and Astrophysics, 57, 467

\bibitem[{R.~D. Blandford \& D. Payne(1982)Blandford \& Payne}]{blandford1982hydromagnetic}
Blandford, R.~D., \& Payne, D. 1982, \bibinfo{title}{Hydromagnetic flows from accretion discs and the production of radio jets,} Monthly Notices of the Royal Astronomical Society, 199, 883

\bibitem[{R.~D. Blandford \& R.~L. Znajek(1977)Blandford \& Znajek}]{blandford1977electromagnetic}
Blandford, R.~D., \& Znajek, R.~L. 1977, \bibinfo{title}{Electromagnetic extraction of energy from Kerr black holes,} Monthly Notices of the Royal Astronomical Society, 179, 433

\bibitem[{C.~J. Burke {et~al.}(2021)Burke, Shen, Blaes, Gammie, Horne, Jiang, Liu, McHardy, Morgan, Scaringi, {et~al.}}]{burke2021characteristic}
Burke, C.~J., Shen, Y., Blaes, O., {et~al.} 2021, \bibinfo{title}{A characteristic optical variability time scale in astrophysical accretion disks,} Science, 373, 789

\bibitem[{M. Carnerero {et~al.}(2017)Carnerero, Raiteri, Villata, Acosta-Pulido, Larionov, Smith, D'Ammando, Agudo, Ar{\'e}valo, Bachev, {et~al.}}]{carnerero2017dissecting}
Carnerero, M., Raiteri, C., Villata, M., {et~al.} 2017, \bibinfo{title}{Dissecting the long-term emission behaviour of the BL Lac object Mrk 421,} Monthly Notices of the Royal Astronomical Society, 472, 3789

\bibitem[{W. Cash(1979)Cash}]{cash1979parameter}
Cash, W. 1979, \bibinfo{title}{Parameter estimation in astronomy through application of the likelihood ratio,} Astrophysical Journal, Part 1, vol. 228, Mar. 15, 1979, p. 939-947., 228, 939

\bibitem[{G. {Castignani} {et~al.}(2013){Castignani}, {Haardt}, {Lapi}, {De Zotti}, {Celotti}, \& {Danese}}]{2013A&A...560A..28C}
{Castignani}, G., {Haardt}, F., {Lapi}, A., {et~al.} 2013, \bibinfo{title}{{Black-hole mass estimates for a homogeneous sample of bright flat-spectrum radio quasars},} \aap, 560, A28, \dodoi{10.1051/0004-6361/201321424}

\bibitem[{B.-Y. Chen {et~al.}(2023)Chen, Bower, Dexter, Markoff, Ridenour, Gurwell, Rao, \& Wallstr{\"o}m}]{chen2023testing}
Chen, B.-Y., Bower, G.~C., Dexter, J., {et~al.} 2023, \bibinfo{title}{Testing the linear relationship between black hole mass and variability timescale in low-luminosity AGNs at submillimeter wavelengths,} The Astrophysical Journal, 951, 93

\bibitem[{K. Cheng {et~al.}(1999)Cheng, Fan, \& Zhang}]{cheng1999basic}
Cheng, K., Fan, J., \& Zhang, L. 1999, \bibinfo{title}{Basic properties of gamma-ray loud blazars,} arXiv preprint astro-ph/9910543

\bibitem[{D.~P. Cohen {et~al.}(2014)Cohen, Romani, Filippenko, Cenko, Lott, Zheng, \& Li}]{cohen2014temporal}
Cohen, D.~P., Romani, R.~W., Filippenko, A.~V., {et~al.} 2014, \bibinfo{title}{Temporal correlations between optical and gamma-ray activity in blazars,} The Astrophysical Journal, 797, 137

\bibitem[{A.~K. Das {et~al.}(2023)Das, Prince, Gupta, \& Kushwaha}]{das2023detection}
Das, A.~K., Prince, R., Gupta, A.~C., \& Kushwaha, P. 2023, \bibinfo{title}{The detection of possible transient quasiperiodic oscillations in the $\gamma$-ray light curve of PKS 0244-470 and 4C+ 38.41,} The Astrophysical Journal, 950, 173

\bibitem[{R. Dixon \& J. Kraus(1968)Dixon \& Kraus}]{dixon1968high}
Dixon, R., \& Kraus, J. 1968, \bibinfo{title}{A High-Sensivity 1415 MHz Survey at North Declinations betwen 19 and 37 degrees.,} Astronomical Journal, Vol. 73, p. 381-407 (1968), 73, 381

\bibitem[{D. Emmanoulopoulos {et~al.}(2013)Emmanoulopoulos, McHardy, \& Papadakis}]{emmanoulopoulos2013generating}
Emmanoulopoulos, D., McHardy, I., \& Papadakis, I. 2013, \bibinfo{title}{Generating artificial light curves: revisited and updated,} Monthly Notices of the Royal Astronomical Society, 433, 907

\bibitem[{J. Fan {et~al.}(1999)Fan, Xie, \& Bacon}]{fan1999central}
Fan, J., Xie, G., \& Bacon, R. 1999, \bibinfo{title}{The central black hole masses and Doppler factors of the $\gamma$-ray loud blazars,} Astronomy and Astrophysics Supplement Series, 136, 13

\bibitem[{J. Fan {et~al.}(2016)Fan, Yang, Liu, Luo, Lin, Yuan, Xiao, Zhou, Hua, \& Pei}]{fan2016spectral}
Fan, J., Yang, J., Liu, Y., {et~al.} 2016, \bibinfo{title}{The spectral energy distributions of Fermi blazars,} The Astrophysical Journal Supplement Series, 226, 20

\bibitem[{J.-H. Fan(2005)Fan}]{fan2005basic}
Fan, J.-H. 2005, \bibinfo{title}{The basic parameters of $\gamma$-ray-loud blazars,} Astronomy \& Astrophysics, 436, 799

\bibitem[{J.-H. Fan {et~al.}(2009)Fan, Yuan, Liu, Zhang, Qin, Liu, Huang, Yang, Wang, \& Zhang}]{fan2009estimations}
Fan, J.-H., Yuan, Y.-H., Liu, Y., {et~al.} 2009, \bibinfo{title}{The estimations of four basic parameters for gamma-ray loud blazars,} Research in Astronomy and Astrophysics, 9, 538

\bibitem[{D. Foreman-Mackey {et~al.}(2013)Foreman-Mackey, Hogg, Lang, \& Goodman}]{foreman2013emcee}
Foreman-Mackey, D., Hogg, D.~W., Lang, D., \& Goodman, J. 2013, \bibinfo{title}{emcee: the MCMC hammer,} Publications of the Astronomical Society of the Pacific, 125, 306

\bibitem[{G.~a. Fossati {et~al.}(1998)Fossati, Maraschi, Celotti, Comastri, \& Ghisellini}]{fossati1998unifying}
Fossati, G.~a., Maraschi, L., Celotti, A., Comastri, A., \& Ghisellini, G. 1998, \bibinfo{title}{A unifying view of the spectral energy distributions of blazars,} Monthly Notices of the Royal Astronomical Society, 299, 433

\bibitem[{G. Foster(1996)Foster}]{foster1996wavelets}
Foster, G. 1996, \bibinfo{title}{Wavelets for period analysis of unevenly sampled time series,} Astronomical Journal v. 112, p. 1709-1729, 112, 1709

\bibitem[{R. Genzel {et~al.}(1997)Genzel, Eckart, Ott, \& Eisenhauer}]{genzel1997nature}
Genzel, R., Eckart, A., Ott, T., \& Eisenhauer, F. 1997, \bibinfo{title}{On the nature of the dark mass in the centre of the Milky Way,} Monthly Notices of the Royal Astronomical Society, 291, 219

\bibitem[{G. Ghisellini \& P. Madau(1996)Ghisellini \& Madau}]{ghisellini1996origin}
Ghisellini, G., \& Madau, P. 1996, \bibinfo{title}{On the origin of the $\gamma$-ray emission in blazars,} Monthly Notices of the Royal Astronomical Society, 280, 67

\bibitem[{G. Ghisellini {et~al.}(1993)Ghisellini, Padovani, Celotti, \& Maraschi}]{ghisellini1993relativistic}
Ghisellini, G., Padovani, P., Celotti, A., \& Maraschi, L. 1993, \bibinfo{title}{Relativistic bulk motion in active galactic nuclei,} Astrophysical Journal, Part 1 (ISSN 0004-637X), vol. 407, no. 1, p. 65-82., 407, 65

\bibitem[{G. {Ghisellini} \& F. {Tavecchio}(2008){Ghisellini} \& {Tavecchio}}]{2008MNRAS.387.1669G}
{Ghisellini}, G., \& {Tavecchio}, F. 2008, \bibinfo{title}{{The blazar sequence: a new perspective},} \mnras, 387, 1669, \dodoi{10.1111/j.1365-2966.2008.13360.x}

\bibitem[{M. Gierli{\'n}ski {et~al.}(2008)Gierli{\'n}ski, Middleton, Ward, \& Done}]{gierlinski2008periodicity}
Gierli{\'n}ski, M., Middleton, M., Ward, M., \& Done, C. 2008, \bibinfo{title}{A periodicity of~ 1 hour in X-ray emission from the active galaxy RE J1034+ 396,} Nature, 455, 369

\bibitem[{P. Giommi {et~al.}(2012)Giommi, Padovani, Polenta, Turriziani, D’Elia, \& Piranomonte}]{giommi2012simplified}
Giommi, P., Padovani, P., Polenta, G., {et~al.} 2012, \bibinfo{title}{A simplified view of blazars: clearing the fog around long-standing selection effects,} Monthly Notices of the Royal Astronomical Society, 420, 2899

\bibitem[{P. Giommi {et~al.}(1999)Giommi, Massaro, Chiappetti, Ferrara, Ghisellini, Jang, Maesano, Miller, Montagni, Nesci, {et~al.}}]{giommi1999synchrotron}
Giommi, P., Massaro, E., Chiappetti, L., {et~al.} 1999, \bibinfo{title}{Synchrotron and inverse Compton variability in the BL Lacertae object S5 0716+ 714,} arXiv preprint astro-ph/9909241

\bibitem[{Y. Gong {et~al.}(2023)Gong, Tian, Zhou, Yi, \& Fang}]{gong2023two}
Gong, Y., Tian, S., Zhou, L., Yi, T., \& Fang, J. 2023, \bibinfo{title}{Two Transient Quasi-periodic Oscillations in $\gamma$-Ray Emission from the Blazar S4 0954+ 658,} The Astrophysical Journal, 949, 39

\bibitem[{Y. Gong {et~al.}(2022{\natexlab{a}})Gong, Zhou, Yuan, Zhang, Yi, \& Fang}]{gong2022quasiperiodic}
Gong, Y., Zhou, L., Yuan, M., {et~al.} 2022{\natexlab{a}}, \bibinfo{title}{Quasiperiodic behavior in the $\gamma$-ray light curve of the blazar PKS 0405-385,} The Astrophysical Journal, 931, 168

\bibitem[{Y. Gong {et~al.}(2022{\natexlab{b}})Gong, Zhou, Yuan, Zhang, Yi, \& Fang}]{Gong_2022}
Gong, Y., Zhou, L., Yuan, M., {et~al.} 2022{\natexlab{b}}, \bibinfo{title}{Quasiperiodic Behavior in the {\ensuremath{\gamma}}-Ray Light Curve of the Blazar PKS 0405-385,} The Astrophysical Journal, 931, 168, \dodoi{10.3847/1538-4357/ac6c8c}

\bibitem[{Y. Gong {et~al.}(2024)Gong, Gao, Li, Yuan, Yi, Li, Qin, Yang, Yang, Zhang, {et~al.}}]{gong2024detection}
Gong, Y., Gao, Q., Li, X., {et~al.} 2024, \bibinfo{title}{The Detection of Possible Quasiperiodic Oscillations in the BL Lac 4FGL J2139. 4- 4235,} The Astrophysical Journal, 976, 51

\bibitem[{M.~J. Graham {et~al.}(2015)Graham, Djorgovski, Stern, Glikman, Drake, Mahabal, Donalek, Larson, \& Christensen}]{graham2015possible}
Graham, M.~J., Djorgovski, S.~G., Stern, D., {et~al.} 2015, \bibinfo{title}{A possible close supermassive black-hole binary in a quasar with optical periodicity,} Nature, 518, 74

\bibitem[{A. Grossmann \& J. Morlet(1984)Grossmann \& Morlet}]{grossmann1984decomposition}
Grossmann, A., \& Morlet, J. 1984, \bibinfo{title}{Decomposition of Hardy functions into square integrable wavelets of constant shape,} SIAM journal on mathematical analysis, 15, 723

\bibitem[{A.~C. Gupta(2017)Gupta}]{gupta2017multi}
Gupta, A.~C. 2017, \bibinfo{title}{Multi-wavelength intra-day variability and quasi-periodic oscillation in blazars,} Galaxies, 6, 1

\bibitem[{A.~C. Gupta {et~al.}(2008{\natexlab{a}})Gupta, Srivastava, \& Wiita}]{gupta2008periodic}
Gupta, A.~C., Srivastava, A., \& Wiita, P.~J. 2008{\natexlab{a}}, \bibinfo{title}{Periodic oscillations in the intra-day optical light curves of the blazar S5 0716+ 714,} The Astrophysical Journal, 690, 216

\bibitem[{A.~C. Gupta {et~al.}(2019{\natexlab{a}})Gupta, Tripathi, Wiita, Kushwaha, Zhang, \& Bambi}]{gupta2019detection}
Gupta, A.~C., Tripathi, A., Wiita, P.~J., {et~al.} 2019{\natexlab{a}}, \bibinfo{title}{Detection of a quasi-periodic oscillation in $\gamma$-ray light curve of the high-redshift blazar B2 1520+ 31,} Monthly Notices of the Royal Astronomical Society, 484, 5785

\bibitem[{A.~C. Gupta {et~al.}(2008{\natexlab{b}})Gupta, Cha, Lee, Jin, Pak, Cho, Moon, Park, Yuk, Nam, {et~al.}}]{gupta2008multicolor}
Gupta, A.~C., Cha, S.-M., Lee, S., {et~al.} 2008{\natexlab{b}}, \bibinfo{title}{Multicolor Near-Infrared Intra-Day and Short-Term Variability of the Blazar S5 0716+ 714,} The Astronomical Journal, 136, 2359

\bibitem[{A.~C. Gupta {et~al.}(2019{\natexlab{b}})Gupta, Gaur, Wiita, Pandey, Kushwaha, Hu, Kurtanidze, Semkov, Damljanovic, Goyal, {et~al.}}]{gupta2019characterizing}
Gupta, A.~C., Gaur, H., Wiita, P.~J., {et~al.} 2019{\natexlab{b}}, \bibinfo{title}{Characterizing Optical Variability of OJ 287 in 2016--2017,} The Astronomical Journal, 157, 95

\bibitem[{Y. Haiyan {et~al.}(2023)Haiyan, Xiefei, Xiaopan, Na, Haitao, Yuhui, Li, \& Yan}]{haiyan2023detection}
Haiyan, Y., Xiefei, S., Xiaopan, L., {et~al.} 2023, \bibinfo{title}{Detection of quasi-periodic oscillation in the optical light curve of the blazar S5 0716+ 714,} Astrophysics and Space Science, 368, 88

\bibitem[{K. Hart {et~al.}(2023)Hart, Shappee, Hey, Kochanek, Stanek, Lim, Dobbs, Tucker, Jayasinghe, Beacom, {et~al.}}]{hart2023asas}
Hart, K., Shappee, B., Hey, D., {et~al.} 2023, \bibinfo{title}{ASAS-SN Sky Patrol V2. 0,} arXiv preprint arXiv:2304.03791

\bibitem[{R. Hartman(1996)Hartman}]{hartman1996gamma}
Hartman, R. 1996, in Blazar continuum variability Astronomical Society of the Pacific Conference Series 110, Proceedings of an international workshop held at Florida International University, Miami, Florida, USA, 4-7 February 1996, San Francisco: Astronomical Society Pacific, edited by H. Richard Miller, James R. Webb, and John C. Noble, p. 333, Vol. 110, 333

\bibitem[{C.-Y. Huang {et~al.}(2013)Huang, Wang, Wang, \& Wang}]{huang2013magnetic}
Huang, C.-Y., Wang, D.-X., Wang, J.-Z., \& Wang, Z.-Y. 2013, \bibinfo{title}{A magnetic reconnection model for quasi-periodic oscillations in black hole systems,} Research in Astronomy and Astrophysics, 13, 705

\bibitem[{S.~G. Jorstad {et~al.}(2022)Jorstad, Marscher, Raiteri, Villata, Weaver, Zhang, Dong, G{\'o}mez, Perel, Savchenko, {et~al.}}]{jorstad2022rapid}
Jorstad, S.~G., Marscher, A., Raiteri, C.~M., {et~al.} 2022, \bibinfo{title}{Rapid quasi-periodic oscillations in the relativistic jet of BL Lacertae,} Nature, 609, 265

\bibitem[{S. Kaspi {et~al.}(2000)Kaspi, Smith, Netzer, Maoz, Jannuzi, \& Giveon}]{kaspi2000reverberation}
Kaspi, S., Smith, P.~S., Netzer, H., {et~al.} 2000, \bibinfo{title}{Reverberation Measurements for 17 Quasars and theSize-Mass-Luminosity Relations in Active GalacticNuclei,} The Astrophysical Journal, 533, 631

\bibitem[{K. Katarzy{\'n}ski {et~al.}(2001)Katarzy{\'n}ski, Sol, \& Kus}]{katarzynski2001multifrequency}
Katarzy{\'n}ski, K., Sol, H., \& Kus, A. 2001, \bibinfo{title}{The multifrequency emission of mrk 501-from radio to tev gamma-rays,} Astronomy \& Astrophysics, 367, 809

\bibitem[{B.~C. Kelly {et~al.}(2009)Kelly, Bechtold, \& Siemiginowska}]{kelly2009variations}
Kelly, B.~C., Bechtold, J., \& Siemiginowska, A. 2009, \bibinfo{title}{Are the variations in quasar optical flux driven by thermal fluctuations?} The Astrophysical Journal, 698, 895

\bibitem[{B.~C. Kelly {et~al.}(2014)Kelly, Becker, Sobolewska, Siemiginowska, \& Uttley}]{kelly2014flexible}
Kelly, B.~C., Becker, A.~C., Sobolewska, M., Siemiginowska, A., \& Uttley, P. 2014, \bibinfo{title}{Flexible and scalable methods for quantifying stochastic variability in the era of massive time-domain astronomical data sets,} The Astrophysical Journal, 788, 33

\bibitem[{P. Kharb {et~al.}(2008)Kharb, Gabuzda, \& Shastri}]{kharb2008parsec}
Kharb, P., Gabuzda, D., \& Shastri, P. 2008, \bibinfo{title}{Parsec-scale magnetic field structures in HEAO-1 BL Lacs,} Monthly Notices of the Royal Astronomical Society, 384, 230

\bibitem[{O. King {et~al.}(2013)King, Hovatta, Max-Moerbeck, Meier, Pearson, Readhead, Reeves, Richards, \& Shepherd}]{king2013quasi}
King, O., Hovatta, T., Max-Moerbeck, W., {et~al.} 2013, \bibinfo{title}{A quasi-periodic oscillation in the blazar J1359+ 4011,} Monthly Notices of the Royal Astronomical Society: Letters, 436, L114

\bibitem[{C. Kochanek {et~al.}(2017)Kochanek, Shappee, Stanek, Holoien, Thompson, Prieto, Dong, Shields, Will, Britt, {et~al.}}]{kochanek2017all}
Kochanek, C., Shappee, B., Stanek, K., {et~al.} 2017, \bibinfo{title}{The all-sky automated survey for supernovae (ASAS-SN) light curve server v1. 0,} Publications of the Astronomical Society of the Pacific, 129, 104502

\bibitem[{S. Koz{\l}owski {et~al.}(2009)Koz{\l}owski, Kochanek, Udalski, Soszy{\'n}ski, Szyma{\'n}ski, Kubiak, Pietrzy{\'n}ski, Szewczyk, Ulaczyk, Poleski, {et~al.}}]{kozlowski2009quantifying}
Koz{\l}owski, S., Kochanek, C.~S., Udalski, A., {et~al.} 2009, \bibinfo{title}{Quantifying quasar variability as part of a general approach to classifying continuously varying sources,} The Astrophysical Journal, 708, 927

\bibitem[{P. Lachowicz {et~al.}(2009)Lachowicz, Gupta, Gaur, \& Wiita}]{lachowicz20094}
Lachowicz, P., Gupta, A., Gaur, H., \& Wiita, P. 2009, \bibinfo{title}{A\~{} 4.6 h quasi-periodic oscillation in the BL Lacertae PKS 2155-304?} Astronomy \& Astrophysics, 506, L17

\bibitem[{H. Li {et~al.}(2016)Li, Jiang, Guo, Chen, \& Yi}]{li2016multiband}
Li, H., Jiang, Y., Guo, D., Chen, X., \& Yi, T. 2016, \bibinfo{title}{Multiband Variability Analysis of Mrk 421,} Publications of the Astronomical Society of the Pacific, 128, 074101

\bibitem[{H.~Z. Li {et~al.}(2015)Li, Chen, Yi, Jiang, Chen, Lü, \& Li}]{Li_2015}
Li, H.~Z., Chen, L.~E., Yi, T.~F., {et~al.} 2015, \bibinfo{title}{Multiband Variability Analysis of 3C 454.3 and Implications for the Center Structure,} Publications of the Astronomical Society of the Pacific, 127, 1, \dodoi{10.1086/679753}

\bibitem[{H.-Z. Li {et~al.}(2024)Li, Qin, Gong, Liu, Guo, Gao, Yi, \& Liu}]{li2024optical}
Li, H.-Z., Qin, L.-H., Gong, Y.-L., {et~al.} 2024, \bibinfo{title}{Optical and $\gamma$-ray variability analysis of BL Lacertae object TXS 1902+ 556,} Monthly Notices of the Royal Astronomical Society, 534, 2986

\bibitem[{H.~Z. Li {et~al.}(2009)Li, Xie, Chen, Dai, Lei, Yi, \& Ren}]{Li_2009}
Li, H.~Z., Xie, G.~Z., Chen, L.~E., {et~al.} 2009, \bibinfo{title}{The Periodicity Analysis of the Light Curve of 3C 279 and Implications for the Precession Jet,} Publications of the Astronomical Society of the Pacific, 121, 1172, \dodoi{10.1086/648433}

\bibitem[{S. Li {et~al.}(2018)Li, Xia, Liang, Liao, \& Fan}]{li2018fast}
Li, S., Xia, Z.-Q., Liang, Y.-F., Liao, N.-H., \& Fan, Y.-Z. 2018, \bibinfo{title}{Fast $\gamma$-Ray Variability in Blazars beyond Redshift 3,} The Astrophysical Journal, 853, 159

\bibitem[{X.-P. Li {et~al.}(2021)Li, Cai, Yang, Luo, Yan, He, \& Wang}]{li2021detection}
Li, X.-P., Cai, Y., Yang, H.-T., {et~al.} 2021, \bibinfo{title}{Detection of quasi-periodic oscillations in the long-term radio light curves of the blazar OT 081,} Monthly Notices of the Royal Astronomical Society, 506, 1540

\bibitem[{X.-P. Li {et~al.}(2023)Li, Cai, Yang, L{\"a}hteenm{\"a}ki, Tornikoski, Tammi, Suutarinen, Yang, Luo, \& Wang}]{li2023quasi}
Li, X.-P., Cai, Y., Yang, H.-Y., {et~al.} 2023, \bibinfo{title}{Quasi-periodic behaviour in the radio and $\gamma$-ray light curves of blazar PKS 1510- 089,} Monthly Notices of the Royal Astronomical Society, 519, 4893

\bibitem[{N. Liao {et~al.}(2014)Liao, Bai, Liu, Weng, Chen, \& Li}]{liao2014multiwavelength}
Liao, N., Bai, J., Liu, H., {et~al.} 2014, \bibinfo{title}{Multiwavelength variability properties of Fermi blazar S5 0716+ 714,} The Astrophysical Journal, 783, 83

\bibitem[{H. Liu \& J. Bai(2015)Liu \& Bai}]{liu2015constraints}
Liu, H., \& Bai, J. 2015, \bibinfo{title}{Constraints on black hole masses with timescales of variations in blazars,} The Astronomical Journal, 149, 191

\bibitem[{N.~R. Lomb(1976)Lomb}]{lomb1976least}
Lomb, N.~R. 1976, \bibinfo{title}{Least-squares frequency analysis of unequally spaced data,} Astrophysics and space science, 39, 447

\bibitem[{L. Lu {et~al.}(2024)Lu, Sun, Fang, Wan, \& Gong}]{lu2024research}
Lu, L., Sun, B., Fang, Z.-X., Wan, M., \& Gong, Y. 2024, \bibinfo{title}{Research on a 44 Day Quasi-periodic Oscillation of Optical Bands for BL Lac S5 0716+ 714,} The Astrophysical Journal, 961, 180

\bibitem[{C.~L. MacLeod {et~al.}(2012)MacLeod, Ivezi{\'c}, Sesar, De~Vries, Kochanek, Kelly, Becker, Lupton, Hall, Richards, {et~al.}}]{macleod2012description}
MacLeod, C.~L., Ivezi{\'c}, {\v{Z}}., Sesar, B., {et~al.} 2012, \bibinfo{title}{A description of quasar variability measured using repeated SDSS and POSS imaging,} The Astrophysical Journal, 753, 106

\bibitem[{M.~P. Malmrose {et~al.}(2011)Malmrose, Marscher, Jorstad, Nikutta, \& Elitzur}]{malmrose2011emission}
Malmrose, M.~P., Marscher, A.~P., Jorstad, S.~G., Nikutta, R., \& Elitzur, M. 2011, \bibinfo{title}{Emission from hot dust in the infrared spectra of gamma-ray bright blazars,} The Astrophysical Journal, 732, 116

\bibitem[{A.~V. Mangalam \& P.~J. Wiita(1993)Mangalam \& Wiita}]{mangalam1993accretion}
Mangalam, A.~V., \& Wiita, P.~J. 1993, \bibinfo{title}{Accretion disk models for optical and ultraviolet microvariability in active galactic nuclei,} Astrophysical Journal, Part 1 (ISSN 0004-637X), vol. 406, no. 2, p. 420-429., 406, 420

\bibitem[{L. Mao \& X. Zhang(2024)Mao \& Zhang}]{mao2024radio}
Mao, L., \& Zhang, X. 2024, \bibinfo{title}{A radio quasi-periodic oscillation in the blazar PKS J2156- 0037,} Monthly Notices of the Royal Astronomical Society, 531, 3927

\bibitem[{L. Maraschi {et~al.}(1992)Maraschi, Ghisellini, \& Celotti}]{maraschi1992jet}
Maraschi, L., Ghisellini, G., \& Celotti, A. 1992, \bibinfo{title}{A jet model for the gamma-ray emitting blazar 3C 279,} Astrophysical Journal, Part 2-Letters (ISSN 0004-637X), vol. 397, no. 1, p. L5-L9., 397, L5

\bibitem[{J.~R. Mattox {et~al.}(1996)Mattox, Bertsch, Chiang, Dingus, Digel, Esposito, Fierro, Hartman, Hunter, Kanbach, {et~al.}}]{mattox1996likelihood}
Mattox, J.~R., Bertsch, D., Chiang, J., {et~al.} 1996, \bibinfo{title}{The likelihood analysis of EGRET data,} Astrophysical Journal v. 461, p. 396, 461, 396

\bibitem[{D.~L. Meier {et~al.}(2001)Meier, Koide, \& Uchida}]{meier2001magnetohydrodynamic}
Meier, D.~L., Koide, S., \& Uchida, Y. 2001, \bibinfo{title}{Magnetohydrodynamic production of relativistic jets,} Science, 291, 84

\bibitem[{P. Mohan \& A. Mangalam(2015)Mohan \& Mangalam}]{mohan2015kinematics}
Mohan, P., \& Mangalam, A. 2015, \bibinfo{title}{Kinematics of and emission from helically orbiting blobs in a relativistic magnetized jet,} The Astrophysical Journal, 805, 91

\bibitem[{J. Moreno {et~al.}(2019)Moreno, Vogeley, Richards, \& Yu}]{moreno2019stochastic}
Moreno, J., Vogeley, M.~S., Richards, G.~T., \& Yu, W. 2019, \bibinfo{title}{Stochastic modeling handbook for optical AGN variability,} Publications of the Astronomical Society of the Pacific, 131, 063001

\bibitem[{J. Otero-Santos {et~al.}(2023)Otero-Santos, Pe{\~n}il, Acosta-Pulido, Becerra~Gonz{\'a}lez, Raiteri, Carnerero, \& Villata}]{otero2023multiwavelength}
Otero-Santos, J., Pe{\~n}il, P., Acosta-Pulido, J., {et~al.} 2023, \bibinfo{title}{Multiwavelength periodicity search in a sample of $\gamma$-ray bright blazars,} Monthly Notices of the Royal Astronomical Society, 518, 5788

\bibitem[{J. Otero-Santos {et~al.}(2020)Otero-Santos, Acosta-Pulido, Becerra~Gonz{\'a}lez, Raiteri, Larionov, Pe{\~n}il, Smith, Ballester~Niebla, Borman, Carnerero, {et~al.}}]{otero2020quasi}
Otero-Santos, J., Acosta-Pulido, J., Becerra~Gonz{\'a}lez, J., {et~al.} 2020, \bibinfo{title}{Quasi-periodic behaviour in the optical and $\gamma$-ray light curves of blazars 3C 66A and B2 1633+ 38,} Monthly Notices of the Royal Astronomical Society, 492, 5524

\bibitem[{P. Padovani(2017)Padovani}]{padovani2017active}
Padovani, P. 2017, \bibinfo{title}{Active Galactic Nuclei at all wavelengths and from all angles,} Frontiers in Astronomy and Space Sciences, 4, 35

\bibitem[{V.~S. Paliya {et~al.}(2021)Paliya, Domínguez, Ajello, Olmo-García, \& Hartmann}]{Paliya_2021}
Paliya, V.~S., Domínguez, A., Ajello, M., Olmo-García, A., \& Hartmann, D. 2021, \bibinfo{title}{The Central Engines of Fermi Blazars,} The Astrophysical Journal Supplement Series, 253, 46, \dodoi{10.3847/1538-4365/abe135}

\bibitem[{H.-W. Pan {et~al.}(2016)Pan, Yuan, Yao, Zhou, Liu, Zhou, \& Zhang}]{pan2016detection}
Pan, H.-W., Yuan, W., Yao, S., {et~al.} 2016, \bibinfo{title}{Detection of a Possible X-ray Quasi-periodic Oscillation in the Active Galactic Nucleus 1H 0707--495,} The Astrophysical Journal Letters, 819, L19

\bibitem[{A. Pandey \& C. Stalin(2022)Pandey \& Stalin}]{pandey2022detection}
Pandey, A., \& Stalin, C. 2022, \bibinfo{title}{Detection of minute-timescale $\gamma$-ray variability in BL Lacertae by Fermi-LAT,} Astronomy \& Astrophysics, 668, A152

\bibitem[{I. Pauliny-Toth {et~al.}(1972)Pauliny-Toth, Kellermann, Davis, Fomalont, \& Shaffer}]{pauliny1972nrao}
Pauliny-Toth, I., Kellermann, K., Davis, M., Fomalont, E., \& Shaffer, D. 1972, \bibinfo{title}{The NRAO 5-GHz radio source survey. II. The 140-ft" strong"," intermediate", and" deep" source surveys.,} Astronomical Journal, Vol. 77, p. 265-284, 77, 265

\bibitem[{Z. Pei {et~al.}(2022)Pei, Fan, Yang, Huang, \& Li}]{pei2022estimation}
Pei, Z., Fan, J., Yang, J., Huang, D., \& Li, Z. 2022, \bibinfo{title}{The Estimation of Fundamental Physics Parameters for Fermi-LAT Blazars,} The Astrophysical Journal, 925, 97

\bibitem[{P. Pe{\~n}il {et~al.}(2020)Pe{\~n}il, Dom{\'\i}nguez, Buson, Ajello, Otero-Santos, Barrio, Nemmen, Cutini, Rani, Franckowiak, {et~al.}}]{penil2020systematic}
Pe{\~n}il, P., Dom{\'\i}nguez, A., Buson, S., {et~al.} 2020, \bibinfo{title}{Systematic search for $\gamma$-ray periodicity in active galactic nuclei detected by the fermi large area telescope,} The Astrophysical Journal, 896, 134

\bibitem[{B.~M. Peterson {et~al.}(1998)Peterson, Wanders, Horne, Collier, Alexander, Kaspi, \& Maoz}]{peterson1998uncertainties}
Peterson, B.~M., Wanders, I., Horne, K., {et~al.} 1998, \bibinfo{title}{On Uncertainties in Cross-Correlation Lags and the Reality of Wavelength-dependent Continuum Lags in Active Galactic Nuclei,} Publications of the Astronomical Society of the Pacific, 110, 660

\bibitem[{B.~M. Peterson {et~al.}(2004)Peterson, Ferrarese, Gilbert, Kaspi, Malkan, Maoz, Merritt, Netzer, Onken, Pogge, {et~al.}}]{peterson2004central}
Peterson, B.~M., Ferrarese, L., Gilbert, K., {et~al.} 2004, \bibinfo{title}{Central masses and broad-line region sizes of active galactic nuclei. II. A homogeneous analysis of a large reverberation-mapping database,} The Astrophysical Journal, 613, 682

\bibitem[{D. Petry {et~al.}(2000)Petry, B{\"o}ttcher, Connaughton, Lahteenmaki, Pursimo, Raiteri, Schr{\"o}der, Sillanp{\"a}{\"a}, Sobrito, Takalo, {et~al.}}]{petry2000multiwavelength}
Petry, D., B{\"o}ttcher, M., Connaughton, V., {et~al.} 2000, \bibinfo{title}{Multiwavelength Observations of Markarian 501 during the 1997 HighState,} The Astrophysical Journal, 536, 742

\bibitem[{R. Prince {et~al.}(2023)Prince, Banerjee, Sharma, Gupta, Bose, {et~al.}}]{prince2023quasi}
Prince, R., Banerjee, A., Sharma, A., {et~al.} 2023, \bibinfo{title}{Quasi-periodic oscillation detected in $\gamma$-rays in blazar PKS 0346- 27,} Astronomy \& Astrophysics, 678, A100

\bibitem[{C.~M. Raiteri {et~al.}(2021{\natexlab{a}})Raiteri, Villata, Carosati, Ben{\'\i}tez, Kurtanidze, Gupta, Mirzaqulov, D’Ammando, Larionov, Pursimo, {et~al.}}]{raiteri2021dual}
Raiteri, C.~M., Villata, M., Carosati, D., {et~al.} 2021{\natexlab{a}}, \bibinfo{title}{The dual nature of blazar fast variability: Space and ground observations of S5 0716+ 714,} Monthly Notices of the Royal Astronomical Society, 501, 1100

\bibitem[{C.~M. Raiteri {et~al.}(2021{\natexlab{b}})Raiteri, Villata, Larionov, Jorstad, Marscher, Weaver, Acosta-Pulido, Agudo, Andreeva, Arkharov, {et~al.}}]{raiteri2021complex}
Raiteri, C.~M., Villata, M., Larionov, V., {et~al.} 2021{\natexlab{b}}, \bibinfo{title}{The complex variability of blazars: Time-scales and periodicity analysis in S4 0954+ 65,} Monthly Notices of the Royal Astronomical Society, 504, 5629

\bibitem[{C. Ren {et~al.}(2024)Ren, Sun, \& Zhang}]{ren2024possible}
Ren, C., Sun, S., \& Zhang, P. 2024, \bibinfo{title}{A Possible Optical Quasiperiodic Oscillation of 134 days in the Radio-loud Narrow-line Seyfert 1 Galaxy TXS 1206+ 549 at z= 1.34,} The Astrophysical Journal, 961, 38

\bibitem[{G.-W. Ren {et~al.}(2021{\natexlab{a}})Ren, Ding, Zhang, Xue, Zhang, Xiong, Li, \& Li}]{ren2021radiodetection}
Ren, G.-W., Ding, N., Zhang, X., {et~al.} 2021{\natexlab{a}}, \bibinfo{title}{Detection of a possible high-confidence radio quasi-periodic oscillation in the BL Lac PKS J2134--0153,} Monthly Notices of the Royal Astronomical Society, 506, 3791

\bibitem[{G.-W. Ren {et~al.}(2021{\natexlab{b}})Ren, Zhang, Zhang, Ding, Yang, Li, Yan, \& Xu}]{ren2021detection}
Ren, G.-W., Zhang, H.-J., Zhang, X., {et~al.} 2021{\natexlab{b}}, \bibinfo{title}{Detection of a high-confidence quasi-periodic oscillation in radio light curve of the high redshift FSRQ PKS J0805--0111,} Research in Astronomy and Astrophysics, 21, 075

\bibitem[{H.~X. Ren {et~al.}(2023)Ren, Cerruti, \& Sahakyan}]{ren2023quasi}
Ren, H.~X., Cerruti, M., \& Sahakyan, N. 2023, \bibinfo{title}{Quasi-periodic oscillations in the $\gamma$-ray light curves of bright active galactic nuclei,} Astronomy \& Astrophysics, 672, A86

\bibitem[{F.~M. Rieger(2004)Rieger}]{rieger2004geometrical}
Rieger, F.~M. 2004, \bibinfo{title}{On the geometrical origin of periodicity in blazar-type sources,} The Astrophysical Journal, 615, L5

\bibitem[{F.~M. Rieger(2005)Rieger}]{rieger2005helical}
Rieger, F.~M. 2005, \bibinfo{title}{Helical Motion and the Origin of QPO in Blazar-type Sources,} Chinese Journal of Astronomy and Astrophysics, 5, 305

\bibitem[{J. {Roland} {et~al.}(2013){Roland}, {Britzen}, {Caproni}, {Fromm}, {Gl{\"u}ck}, \& {Zensus}}]{2013A&A...557A..85R}
{Roland}, J., {Britzen}, S., {Caproni}, A., {et~al.} 2013, \bibinfo{title}{{Binary black holes in nuclei of extragalactic radio sources},} \aap, 557, A85, \dodoi{10.1051/0004-6361/201219165}

\bibitem[{G.~E. Romero {et~al.}(2000)Romero, Chajet, Abraham, \& Fan}]{romero2000beaming}
Romero, G.~E., Chajet, L.~S., Abraham, Z., \& Fan, J.~H. 2000, \bibinfo{title}{Beaming and precession in the inner jet of 3C 273,} Astronomy and Astrophysics, 360

\bibitem[{A. Roy {et~al.}(2022{\natexlab{a}})Roy, Sarkar, Chatterjee, Gupta, Chitnis, \& Wiita}]{roy2022transient}
Roy, A., Sarkar, A., Chatterjee, A., {et~al.} 2022{\natexlab{a}}, \bibinfo{title}{Transient quasi-periodic oscillations at $\gamma$-rays in the TeV blazar PKS 1510-089,} Monthly Notices of the Royal Astronomical Society, 510, 3641

\bibitem[{A. Roy {et~al.}(2022{\natexlab{b}})Roy, Chitnis, Gupta, Wiita, Romero, Cellone, Chatterjee, Combi, Raiteri, Sarkar, {et~al.}}]{roy2022detection}
Roy, A., Chitnis, V.~R., Gupta, A.~C., {et~al.} 2022{\natexlab{b}}, \bibinfo{title}{Detection of a quasi-periodic oscillation in the optical light curve of the remarkable blazar AO 0235+ 164,} Monthly Notices of the Royal Astronomical Society, 513, 5238

\bibitem[{J.~J. Ruan {et~al.}(2012)Ruan, Anderson, MacLeod, Becker, Burnett, Davenport, Ivezi{\'c}, Kochanek, Plotkin, Sesar, {et~al.}}]{ruan2012characterizing}
Ruan, J.~J., Anderson, S.~F., MacLeod, C.~L., {et~al.} 2012, \bibinfo{title}{Characterizing the optical variability of bright blazars: Variability-based selection of fermi active galactic nuclei,} The Astrophysical Journal, 760, 51

\bibitem[{J.~L. Ryan {et~al.}(2019)Ryan, Siemiginowska, Sobolewska, \& Grindlay}]{ryan2019characteristic}
Ryan, J.~L., Siemiginowska, A., Sobolewska, M., \& Grindlay, J. 2019, \bibinfo{title}{Characteristic variability timescales in the gamma-ray power spectra of blazars,} The Astrophysical Journal, 885, 12

\bibitem[{A. Sandrinelli {et~al.}(2016)Sandrinelli, Covino, Dotti, \& Treves}]{sandrinelli2016quasi}
Sandrinelli, A., Covino, S., Dotti, M., \& Treves, A. 2016, \bibinfo{title}{Quasi-periodicities at Year-like Timescales in Blazars,} The Astronomical Journal, 151, 54

\bibitem[{A. Sandrinelli {et~al.}(2014)Sandrinelli, Covino, \& Treves}]{sandrinelli2014long}
Sandrinelli, A., Covino, S., \& Treves, A. 2014, \bibinfo{title}{Long and short term variability of seven blazars in six near-infrared/optical bands,} Astronomy \& Astrophysics, 562, A79

\bibitem[{A. Sarkar {et~al.}(2021)Sarkar, Gupta, Chitnis, \& Wiita}]{sarkar2021multiwaveband}
Sarkar, A., Gupta, A.~C., Chitnis, V.~R., \& Wiita, P.~J. 2021, \bibinfo{title}{Multiwaveband quasi-periodic oscillation in the blazar 3C 454.3,} Monthly Notices of the Royal Astronomical Society, 501, 50

\bibitem[{A. Sarkar {et~al.}(2019)Sarkar, Chitnis, Gupta, Gaur, Patel, Wiita, Volvach, Tornikoski, Chamani, Enestam, {et~al.}}]{sarkar2019long}
Sarkar, A., Chitnis, V., Gupta, A., {et~al.} 2019, \bibinfo{title}{Long-term variability and correlation study of the blazar 3C 454.3 in the radio, NIR, and optical wavebands,} The Astrophysical Journal, 887, 185

\bibitem[{J.~D. Scargle(1979)Scargle}]{scargle1979studies}
Scargle, J.~D. 1979, Studies in astronomical time series analysis: Modeling random processes in the time domain, Vol. 81148 (National Aeronautics and Space Administration, Ames Research Center)

\bibitem[{M. Schulz \& M. Mudelsee(2002)Schulz \& Mudelsee}]{schulz2002redfit}
Schulz, M., \& Mudelsee, M. 2002, \bibinfo{title}{REDFIT: estimating red-noise spectra directly from unevenly spaced paleoclimatic time series,} Computers \& Geosciences, 28, 421

\bibitem[{B.~J. Shappee {et~al.}(2014)Shappee, Prieto, Grupe, Kochanek, Stanek, De~Rosa, Mathur, Zu, Peterson, Pogge, {et~al.}}]{shappee2014man}
Shappee, B.~J., Prieto, J., Grupe, D., {et~al.} 2014, \bibinfo{title}{The man behind the curtain: X-rays drive the UV through NIR variability in the 2013 active galactic nucleus outburst in NGC 2617,} The Astrophysical Journal, 788, 48

\bibitem[{A. Sharma {et~al.}(2024{\natexlab{a}})Sharma, Banerjee, Das, Mandal, \& Bose}]{sharma2024detection}
Sharma, A., Banerjee, A., Das, A.~K., Mandal, A., \& Bose, D. 2024{\natexlab{a}}, \bibinfo{title}{Detection of a Transient Quasiperiodic Oscillation in $\gamma$-Rays from Blazar PKS 2255-282,} The Astrophysical Journal, 975, 56

\bibitem[{A. Sharma {et~al.}(2024{\natexlab{b}})Sharma, Kamaram, Prince, Khatoon, \& Bose}]{sharma2024probing}
Sharma, A., Kamaram, S.~R., Prince, R., Khatoon, R., \& Bose, D. 2024{\natexlab{b}}, \bibinfo{title}{Probing the disc--jet coupling in S4 0954+ 65, PKS 0903- 57, and 4C+ 01.02 with $\gamma$-rays,} Monthly Notices of the Royal Astronomical Society, 527, 2672

\bibitem[{A. Sharma {et~al.}(2024{\natexlab{c}})Sharma, Prince, \& Bose}]{sharma2024microquasars}
Sharma, A., Prince, R., \& Bose, D. 2024{\natexlab{c}}, \bibinfo{title}{Microquasars to AGNs: An uniform Jet variability,} arXiv preprint arXiv:2410.06653

\bibitem[{A. Sillanpaa {et~al.}(1988)Sillanpaa, Haarala, Valtonen, Sundelius, \& Byrd}]{sillanpaa1988oj}
Sillanpaa, A., Haarala, S., Valtonen, M., Sundelius, B., \& Byrd, G. 1988, \bibinfo{title}{OJ 287-Binary pair of supermassive black holes,} Astrophysical Journal, Part 1 (ISSN 0004-637X), vol. 325, Feb. 15, 1988, p. 628-634. Research supported by Nordisk Institut for Teoretisk Atomfysik and NSF., 325, 628

\bibitem[{E. Smith {et~al.}(2023)Smith, Oramas, \& Perlman}]{smith2023qpo}
Smith, E., Oramas, L., \& Perlman, E. 2023, \bibinfo{title}{A QPO in Mkn 421 from Archival RXTE Data,} The Astrophysical Journal, 950, 174

\bibitem[{E. {Sobacchi} {et~al.}(2017){Sobacchi}, {Sormani}, \& {Stamerra}}]{2017MNRAS.465..161S}
{Sobacchi}, E., {Sormani}, M.~C., \& {Stamerra}, A. 2017, \bibinfo{title}{{A model for periodic blazars},} \mnras, 465, 161, \dodoi{10.1093/mnras/stw2684}

\bibitem[{M.~A. Sobolewska {et~al.}(2014)Sobolewska, Siemiginowska, Kelly, \& Nalewajko}]{sobolewska2014stochastic}
Sobolewska, M.~A., Siemiginowska, A., Kelly, B.~C., \& Nalewajko, K. 2014, \bibinfo{title}{Stochastic modeling of the Fermi/LAT $\gamma$-ray blazar variability,} The Astrophysical Journal, 786, 143

\bibitem[{A. So{\l}tan(1982)So{\l}tan}]{soƚtan1982masses}
So{\l}tan, A. 1982, \bibinfo{title}{Masses of quasars,} Monthly Notices of the Royal Astronomical Society, 200, 115

\bibitem[{L. Stella \& M. Vietri(1997)Stella \& Vietri}]{stella1997lense}
Stella, L., \& Vietri, M. 1997, \bibinfo{title}{Lense-Thirring precession and quasi-periodic oscillations in low-mass X-ray binaries,} The Astrophysical Journal, 492, L59

\bibitem[{M. {Sun} {et~al.}(2018){Sun}, {Grier}, \& {Peterson}}]{2018ascl.soft05032S}
{Sun}, M., {Grier}, C.~J., \& {Peterson}, B.~M. 2018, \bibinfo{title}{{PyCCF: Python Cross Correlation Function for reverberation mapping studies},}, Astrophysics Source Code Library, record ascl:1805.032

\bibitem[{G. Tagliaferri {et~al.}(2003)Tagliaferri, Ravasio, Ghisellini, Tavecchio, Giommi, Massaro, Nesci, \& Tosti}]{tagliaferri2003wide}
Tagliaferri, G., Ravasio, M., Ghisellini, G., {et~al.} 2003, \bibinfo{title}{Wide band X-ray and optical observations of the BL Lac object 1ES 1959+ 650 in high state,} Astronomy \& Astrophysics, 412, 711

\bibitem[{J. Tantry {et~al.}(2025)Tantry, Sharma, Shah, Iqbal, \& Bose}]{tantry2025study}
Tantry, J., Sharma, A., Shah, Z., Iqbal, N., \& Bose, D. 2025, \bibinfo{title}{Study of multi-wavelength variability, emission mechanism and quasi-periodic oscillation for transition blazar S5 1803+ 784,} Journal of High Energy Astrophysics, 100372

\bibitem[{M. Tavani {et~al.}(2018)Tavani, Cavaliere, Munar-Adrover, \& Argan}]{tavani2018blazar}
Tavani, M., Cavaliere, A., Munar-Adrover, P., \& Argan, A. 2018, \bibinfo{title}{The blazar PG 1553+ 113 as a binary system of supermassive black holes,} The Astrophysical Journal, 854, 11

\bibitem[{F. Tavecchio {et~al.}(1998)Tavecchio, Maraschi, \& Ghisellini}]{tavecchio1998constraints}
Tavecchio, F., Maraschi, L., \& Ghisellini, G. 1998, \bibinfo{title}{Constraints on the physical parameters of TeV blazars,} The Astrophysical Journal, 509, 608

\bibitem[{A. Tripathi {et~al.}(2021)Tripathi, Gupta, Aller, Wiita, Bambi, Aller, \& Gu}]{tripathi2021quasi}
Tripathi, A., Gupta, A.~C., Aller, M.~F., {et~al.} 2021, \bibinfo{title}{Quasi-periodic oscillations in the long-term radio light curves of the blazar AO 0235+ 164,} Monthly Notices of the Royal Astronomical Society, 501, 5997

\bibitem[{M.-H. Ulrich {et~al.}(1997)Ulrich, Maraschi, \& Urry}]{ulrich1997variability}
Ulrich, M.-H., Maraschi, L., \& Urry, C.~M. 1997, \bibinfo{title}{Variability of active galactic nuclei,} Annual Review of Astronomy and Astrophysics, 35, 445

\bibitem[{C. Urry {et~al.}(1993)Urry, Maraschi, Edelson, Koratkar, Krolik, Madejski, Pian, Pike, Reichert, Treves, {et~al.}}]{urry1993multiwavelength}
Urry, C., Maraschi, L., Edelson, R., {et~al.} 1993, \bibinfo{title}{Multiwavelength monitoring of the BL Lacertae object PKS 2155-304. I-The IUE campaign,} Astrophysical Journal-Part 1 (ISSN 0004-637X), vol. 411, no. 2, p. 614-631., 411, 614

\bibitem[{C.~M. Urry \& P. Padovani(1995)Urry \& Padovani}]{urry1995unified}
Urry, C.~M., \& Padovani, P. 1995, \bibinfo{title}{Unified schemes for radio-loud active galactic nuclei,} Publications of the Astronomical Society of the Pacific, 107, 803

\bibitem[{C.~M. {Urry} \& P. {Padovani}(1995){Urry} \& {Padovani}}]{1995PASP..107..803U}
{Urry}, C.~M., \& {Padovani}, P. 1995, \bibinfo{title}{{Unified Schemes for Radio-Loud Active Galactic Nuclei},} \pasp, 107, 803, \dodoi{10.1086/133630}

\bibitem[{M. Valtonen {et~al.}(2011)Valtonen, Lehto, Takalo, \& Sillanp{\"a}{\"a}}]{valtonen2011testing}
Valtonen, M., Lehto, H., Takalo, L., \& Sillanp{\"a}{\"a}, A. 2011, \bibinfo{title}{Testing the 1995 binary black hole model of OJ287,} The Astrophysical Journal, 729, 33

\bibitem[{M.~J. Valtonen {et~al.}(2008)Valtonen, Lehto, Nilsson, Heidt, Takalo, Sillanp{\"a}{\"a}, Villforth, Kidger, Poyner, Pursimo, {et~al.}}]{valtonen2008massive}
Valtonen, M.~J., Lehto, H., Nilsson, K., {et~al.} 2008, \bibinfo{title}{A massive binary black-hole system in OJ 287 and a test of general relativity,} Nature, 452, 851

\bibitem[{J.~T. VanderPlas(2018)VanderPlas}]{vanderplas2018understanding}
VanderPlas, J.~T. 2018, \bibinfo{title}{Understanding the lomb--scargle periodogram,} The Astrophysical Journal Supplement Series, 236, 16

\bibitem[{M. Vestergaard(2002)Vestergaard}]{vestergaard2002determining}
Vestergaard, M. 2002, \bibinfo{title}{Determining central black hole masses in distant active galaxies,} The Astrophysical Journal, 571, 733

\bibitem[{M. Villata \& C. Raiteri(1999)Villata \& Raiteri}]{villata1999helical}
Villata, M., \& Raiteri, C. 1999, \bibinfo{title}{Helical jets in blazars. I. The case of MKN 501,} Astronomy and Astrophysics, v. 347, p. 30-36 (1999), 347, 30

\bibitem[{C. Villforth {et~al.}(2010)Villforth, Nilsson, Heidt, Takalo, Pursimo, Berdyugin, Lindfors, Pasanen, Winiarski, Drozdz, {et~al.}}]{villforth2010variability}
Villforth, C., Nilsson, K., Heidt, J., {et~al.} 2010, \bibinfo{title}{Variability and stability in blazar jets on time-scales of years: Optical polarization monitoring of OJ 287 in 2005--2009,} Monthly Notices of the Royal Astronomical Society, 402, 2087

\bibitem[{S. Wagner \& A. Witzel(1995)Wagner \& Witzel}]{wagner1995intraday}
Wagner, S., \& Witzel, A. 1995, \bibinfo{title}{Intraday variability in quasars and BL Lac objects,} Annual Review of Astronomy and Astrophysics, 33, 163

\bibitem[{G. Wang {et~al.}(2022)Wang, Cai, \& Fan}]{wang2022possible}
Wang, G., Cai, J., \& Fan, J. 2022, \bibinfo{title}{A Possible 3 yr Quasi-periodic Oscillation in $\gamma$-Ray Emission from the FSRQ S5 1044+ 71,} The Astrophysical Journal, 929, 130

\bibitem[{H. Wang {et~al.}(2017)Wang, Yin, \& Xiang}]{wang2017nearly}
Wang, H., Yin, C., \& Xiang, F. 2017, \bibinfo{title}{The nearly periodic fluctuations of blazars in long-term X-ray light curves,} Astrophysics and Space Science, 362, 1

\bibitem[{J.-Y. Wang {et~al.}(2014)Wang, An, Baan, \& Lu}]{10.1093/mnras/stu1135}
Wang, J.-Y., An, T., Baan, W.~A., \& Lu, X.-L. 2014, \bibinfo{title}{Periodic radio variabilities of the blazar 1156+295: harmonic oscillations,} Monthly Notices of the Royal Astronomical Society, 443, 58, \dodoi{10.1093/mnras/stu1135}

\bibitem[{Z.-Z. Wu {et~al.}(2009)Wu, Gu, \& Jiang}]{wu2009debeamed}
Wu, Z.-Z., Gu, M.-F., \& Jiang, D.-R. 2009, \bibinfo{title}{The debeamed luminosity, sychrotron peak frequency and black hole mass of BL Lac objects,} Research in Astronomy and Astrophysics, 9, 168

\bibitem[{G. Xie {et~al.}(1998)Xie, Bai, Zhang, \& Fan}]{xie1998massive}
Xie, G., Bai, J., Zhang, X., \& Fan, J. 1998, \bibinfo{title}{The massive black hole in the center of the active galaxy MRK 421,} Astronomy and Astrophysics, v. 334, p. L29-L31 (1998), 334, L29

\bibitem[{G. Xie {et~al.}(2008)Xie, Yi, Li, Zhou, \& Chen}]{xie2008periodicity}
Xie, G., Yi, T., Li, H., Zhou, S., \& Chen, L. 2008, \bibinfo{title}{Periodicity analysis of the radio curve of PKS 1510-089 and implications for its central structure,} The Astronomical Journal, 135, 2212

\bibitem[{D.~R. Xiong \& X. Zhang(2014)Xiong \& Zhang}]{10.1093/mnras/stu755}
Xiong, D.~R., \& Zhang, X. 2014, \bibinfo{title}{Intrinsic $\gamma$-ray luminosity, black hole mass, jet and accretion in Fermi blazars,} Monthly Notices of the Royal Astronomical Society, 441, 3375, \dodoi{10.1093/mnras/stu755}

\bibitem[{J. {Yang} {et~al.}(2021){Yang}, {Cao}, {Zhou}, \& {Qin}}]{2021PASP..133b4101Y}
{Yang}, J., {Cao}, G., {Zhou}, B., \& {Qin}, L. 2021, \bibinfo{title}{{Quasi-periodic Oscillation of Blazar PKS 1424-418 in {\ensuremath{\gamma}}-Ray Band},} \pasp, 133, 024101, \dodoi{10.1088/1538-3873/abd152}

\bibitem[{J. Yang \& J. Fan(2010)Yang \& Fan}]{yang2010central}
Yang, J., \& Fan, J. 2010, \bibinfo{title}{The central black hole masses for the $\gamma$-ray loud blazars,} Science China Physics, Mechanics and Astronomy, 53, 1921

\bibitem[{S. Yang {et~al.}(2021)Yang, Yan, Zhang, Dai, \& Zhang}]{yang2021gaussian}
Yang, S., Yan, D., Zhang, P., Dai, B., \& Zhang, L. 2021, \bibinfo{title}{Gaussian process modeling fermi-lat $\gamma$-ray blazar variability: A sample of blazars with $\gamma$-ray quasi-periodicities,} The Astrophysical Journal, 907, 105

\bibitem[{H. Zhang {et~al.}(2022)Zhang, Yan, \& Zhang}]{zhang2022characterizing}
Zhang, H., Yan, D., \& Zhang, L. 2022, \bibinfo{title}{Characterizing the $\gamma$-ray variability of active galactic nuclei with the stochastic process method,} The Astrophysical Journal, 930, 157

\bibitem[{H. Zhang {et~al.}(2023)Zhang, Yan, \& Zhang}]{zhang2023gaussian}
Zhang, H., Yan, D., \& Zhang, L. 2023, \bibinfo{title}{Gaussian Process Modeling Blazar Multiwavelength Variability: Indirectly Resolving Jet Structure,} The Astrophysical Journal, 944, 103

\bibitem[{H. Zhang {et~al.}(2024)Zhang, Yang, \& Dai}]{zhang2024discovering}
Zhang, H., Yang, S., \& Dai, B. 2024, \bibinfo{title}{Discovering the Mass-Scaled Damping Timescale from Microquasars to Blazars,} The Astrophysical Journal Letters, 967, L18

\bibitem[{P.-F. Zhang {et~al.}(2017)Zhang, Yan, Zhou, Fan, Wang, \& Zhang}]{zhang2017gamma}
Zhang, P.-F., Yan, D.-H., Zhou, J.-N., {et~al.} 2017, \bibinfo{title}{A $\gamma$-ray Quasi-periodic Modulation in the Blazar PKS 0301--243?} The Astrophysical Journal, 845, 82

\bibitem[{X. Zhang \& G. Bao(1990)Zhang \& Bao}]{zhang1990rotation}
Zhang, X., \& Bao, G. 1990, \bibinfo{title}{The rotation of accretion-disks and the power spectra of X-rays' flickering',}, Tech. rep., International Centre for Theoretical Physics

\bibitem[{X. Zhang {et~al.}(2024)Zhang, Xiong, Gao, Yang, Lu, Na, \& Qin}]{10.1093/mnras/stae765}
Zhang, X., Xiong, D.-r., Gao, Q.-g., {et~al.} 2024, \bibinfo{title}{The fundamental plane of blazars based on the black hole spin-mass energy,} Monthly Notices of the Royal Astronomical Society, 529, 3699, \dodoi{10.1093/mnras/stae765}

\bibitem[{Y. Zheng {et~al.}(2013)Zheng, Zhang, Huang, \& Kang}]{zheng2013modelling}
Zheng, Y., Zhang, L., Huang, B., \& Kang, S. 2013, \bibinfo{title}{Modelling the $\gamma$-ray variability of 3C 273,} Monthly Notices of the Royal Astronomical Society, 431, 2356

\bibitem[{J. Zhou {et~al.}(2018)Zhou, Wang, Chen, Wiita, Vadakkumthani, Morrell, Zhang, \& Zhang}]{zhou201834}
Zhou, J., Wang, Z., Chen, L., {et~al.} 2018, \bibinfo{title}{A 34.5 day quasi-periodic oscillation in $\gamma$-ray emission from the blazar PKS 2247--131,} Nature communications, 9, 4599

\bibitem[{R. Zhou {et~al.}(2021)Zhou, Zheng, Zhu, \& Kang}]{zhou2021intrinsic}
Zhou, R., Zheng, Y., Zhu, K., \& Kang, S. 2021, \bibinfo{title}{The Intrinsic Properties of Multiwavelength Energy Spectra for Fermi Teraelectronvolt Blazars,} The Astrophysical Journal, 915, 59

\bibitem[{Y. Zu {et~al.}(2013)Zu, Kochanek, Koz{\l}owski, \& Udalski}]{zu2013quasar}
Zu, Y., Kochanek, C., Koz{\l}owski, S., \& Udalski, A. 2013, \bibinfo{title}{Is quasar optical variability a damped random walk?} The Astrophysical Journal, 765, 106

\end{thebibliography}
\bibliographystyle{aasjournalv7}



\end{document}